\documentclass[aps,pra,superscriptaddress,notitlepage,twocolumn,floatfix]{revtex4-2}
\pdfoutput=1 % Make sure it compiles on arxiv
\usepackage[utf8]{inputenc}
\usepackage{amsmath,amsthm,amssymb,amsfonts,physics,bbm}
\usepackage{graphicx}
\usepackage{xcolor}

\DeclareMathOperator{\sgn}{sgn}
\DeclareMathOperator{\pos}{pos}
\allowdisplaybreaks

\usepackage{hyperref}
\hypersetup{
    colorlinks = true,
    citecolor = blue,
    linkcolor = blue
}

\begin{document}

\title{Dynamics-based quantumness certification of continuous variables using time-independent Hamiltonians with one degree of freedom}
\author{Lin Htoo Zaw}
\affiliation{Centre for Quantum Technologies, National University of Singapore, 3 Science Drive 2, Singapore 117543}
\author{Valerio Scarani}
\affiliation{Centre for Quantum Technologies, National University of Singapore, 3 Science Drive 2, Singapore 117543}
\affiliation{Department of Physics, National University of Singapore, 2 Science Drive 3, Singapore 117542}
%\date{}

\begin{abstract}
Dynamics-based certification of quantumness is an approach to witnessing the nonclassical character of some continuous-variable states, under the assumption that their dynamics is known. Contrary to other tests of nonclassicality for single systems, it does not require sequential measurements. This family of protocols was introduced for harmonic dynamics. In this work, we discuss dynamics-based certification for one degree of freedom evolving under a generic time-independent Hamiltonian. We characterize the conditions under which such a certification is possible. Several examples are explicitly studied: some that are approximately harmonic in the limits of low energy (Kerr nonlinearities, the pendulum, and the Morse potential) and one that is not (the particle in an infinite well).
\end{abstract}

\maketitle

\section{Introduction}

The demonstration of quantum effects with large objects is a current frontier of research. Demonstrations of typical wave effects like interference are compelling, but are not the only options: Nonclassical features within the mechanical framework are certainly more suited for trapped systems. The most famous nonclassical features appear as negatives compared to classical theory: For instance, a material point's position and momentum cannot be localized beyond the limits of Heisenberg's uncertainty. The detection of such negative features relies on calibration: Only by measuring at a precision dictated by Planck's constant can one claim that the remaining fluctuations are unavoidable (and the claim would not be accepted by die-hard classical physicists, assuming there are still some of those).

In this paper, we propose a family of positive criteria of nonclassicality, i.e.~observations that are impossible to obtain with any classical state. Our criteria are dynamics-based (they assume that the time evolution of the system is known) and state-dependent (only certain states, usually with suitable negativity in their Wigner function, can be detected). Contrary to tests based on contextuality arguments or the Leggett-Garg approach, our tests do not depend on sequential or simultaneous measurements, instead requiring only one measurement on each system for every round of the test.

Given a Hamiltonian of the form $H(x,p)={p^2}/{2m}+V(x)$, Ehrenfest's theorem proves that the Heisenberg equations of motion are the same as Hamilton's equations, with the canonical variables replaced by the corresponding pair of quantum observables that satisfy the canonical commutation relations. The solutions for position and momentum as a function of time have the same expressions for classical variables and quantum observables. Remarkably, the first examples of dynamics-based nonclassicality criteria were built on such ostensibly classical dynamics. An early example is the probability backflow of a free particle whose momentum is always found to be positive \cite{Backflow}, which was recently extended in the context of the demonstration of quantum advantage in mechanical tasks \cite{Rocket}. By these criteria, nonclassical behavior is found by measuring the position of the system after waiting for a set duration. Similar signatures of quantumness can be found in the time evolution of the harmonic oscillator, which is a simple precession in phase space in both the classical and quantum cases. Tsirelson introduced a nonclassicality test that involves measuring the coordinate of the system once per round, at a randomly chosen time each round \cite{tsirelson}. Extending this idea, a family of protocols was recently introduced for detecting the nonclassicality of a single quantum system under the sole assumption that its dynamics is a uniform precession \cite{Zaw2022}. When applied to a normal mode of coupled oscillators, this criterion becomes an entanglement witness \cite{pooja}.

These dynamics-based tests of nonclassicality utilize coarse-grained measurements of continuous variables like $\sgn(x)$. Such dichotomic measurements have also been used for nonlocality tests for two continuous-variable systems \cite{Bell-sgnX,Bell-parity,Bell-fakeSpin} and for single-system nonclassicality tests which are based on some other assumptions rather than the dynamics \cite{LG-original-sgnX,Wronskian-trick,Wronskian-trick-followup,LG-Massive}.

\begin{figure*}
    \includegraphics[width=\textwidth]{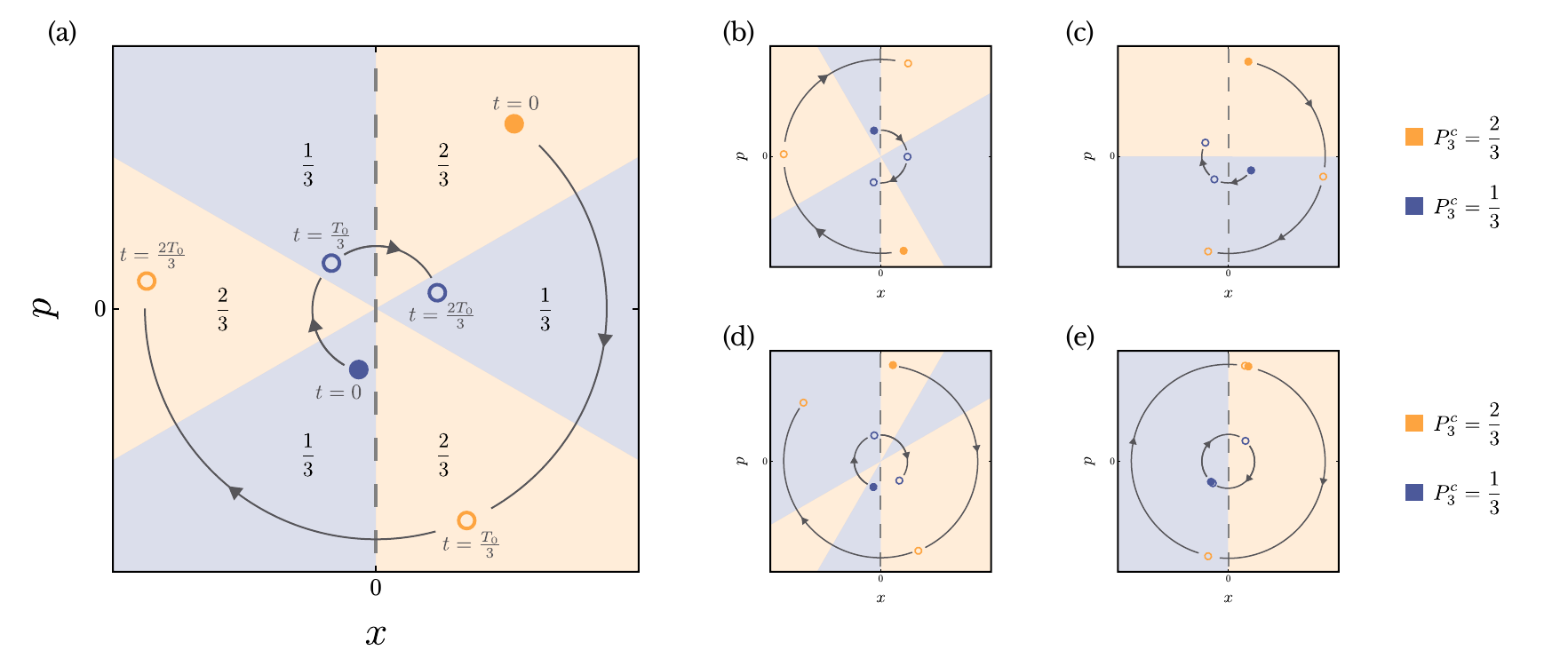}
    \caption{\label{fig:Harmonic_initial_states}For the harmonic oscillator, in classical theory, any pure state $(x(t=0),p(t=0))$ initially in the orange region achieves the score of $P_3^c = \frac{2}{3}$, while any pure state initially in the blue region achieves the score of $P_3^c = \frac{1}{3}$. Some example trajectories are shown for different probing duration $\mathbf{T}$, with the closed points marking its state at $t=0$ and the open points marking its states at $t=\mathbf{T}/3$ and $2\mathbf{T}/3$. Consider first (a), with $\mathbf{T}=T_0$. A state initially in the top-right orange region will evolve to the bottom-right orange region at $t=\mathbf{T}/3$ and then the leftmost orange region at $t=2\mathbf{T}/3$. Therefore, that state will achieve the score $P_3^c=\frac{2}{3}$. Repeating the same argument for every other state initially in any of the orange regions would lead us to conclude that any state initially in any of the orange regions will have $P_3^c=\frac{2}{3}$. Similarly, a state initially in the bottom-left blue region will evolve to the top-left blue region at $t=\mathbf{T}/3$ and rightmost blue region at $t=2\mathbf{T}/3$. The same argument for every other state initially in any of the blue regions will have $P_3^c=\frac{1}{3}$. Analogous arguments hold for different probing duration $\mathbf{T}$, with decreasing $\mathbf{T}$ from left to right with (b) $3T_0/4 < \mathbf{T} < T_0$ and (c) $\mathbf{T} = 3T_0/4$, and increasing $\mathbf{T}$ from left to right with (d) $T_0 < \mathbf{T} < 3T_0/2$ and (e) $\mathbf{T} = 3T_0/2$. Other intermediate cases for $3T_0/4 < \mathbf{T} < T_0$ and $T_0 < \mathbf{T} < 3T_0/2$ are similar to (b) and (d), respectively. }
\end{figure*}

In the original criterion and its extension \cite{tsirelson,Zaw2022}, the primary assumption was that the time evolution of the system is a uniform precession. The contribution of this paper is to extend dynamics-based nonclassicality criteria to a conserved (i.e.,~time-independent) Hamiltonian of one continuous degree of freedom. A quantum-classical gap is not always present: For our criterion to certify nonclassicality, the dynamics should lead to some form of trapping (precise conditions in Sec.~\ref{sec:general}). While in most examples, and probably in all tabletop implementations, trapping potentials are approximately harmonic at low energy, this will not be a requirement for our protocol to detect nonclassicality.

After presenting the general approach, we apply it to several examples, which go beyond the original harmonic case with nontrivial variations. The first is the Kerr anharmonic Hamiltonian (Sec.~\ref{sec:example-kerr}), which contains quartic terms and describes several real quantum devices with nonlinearities. Second, we study the pendulum (Sec.~\ref{sec:example-pendulum}), which is a Hamiltonian in angular, rather than linear, coordinates. Third, we introduce the Morse potential, a potential that is asymmetric and not fully trapping, i.e.,~has open orbits at high energy (Sec.~\ref{sec:example-morse}). Finally, we study the infinite potential well (Sec.~\ref{sec:example-infinite-well}), which is not harmonic even in the limit of small energy; this shows that our protocol is not based on approximate harmonicity.

For each of these examples, we find the range of parameters for which nonclassicality can be detected with this protocol. For the approximately harmonic potentials, we also compare the complete dynamics against the harmonic approximation in the low-energy regime. Rather than order-of-magnitude approximations, we are able to give exact values of anharmonicities where the nonclassicality criterion stops working. Furthermore, we also show that a simple adjustment in the probing time permits a violation even at larger anharmonicities.

Before discussing the new potentials, we revisit the study of the harmonic oscillator in Sec.~\ref{sec:revisit}, introducing the original protocol and some variations on it.

\section{Revisiting the protocol for the harmonic oscillator}
\label{sec:revisit}

To better understand the formulation of the general criterion, it is useful to revisit the known and simple case of the harmonic oscillator $H(x,p) = p^2/{2m} + m\omega_0^2 x^2/{2}$. It is well known that the dynamics is such that $(x(t),p(t))$ undergoes a uniform precession in phase space with the period $T_0=2\pi/\omega_0$. In particular, $x(t) = x(0)\cos\omega_0 t + [p(0)/m\omega_0] \sin\omega_0 t$.

\subsection{Overview of the original protocol\texorpdfstring{ \cite{tsirelson,Zaw2022}}{}}\label{sec:original-steps}

The original protocol consists of many independent rounds. The following steps take place in in each round.

(1) One system is prepared in some state. For the validity of the certification, one does not need to assume that the state is the same in every round (of course, poor preparation will lead to a negative outcome of the certification).

(2) After the preparation is completed, the system is decoupled from everything else and undergoes the closed dynamics of a harmonic oscillator, resulting in a uniform precession with period $T_0=2\pi/\omega_0$.

(3) A duration $t \in \{ 0, T_0/3, 2T_0/3 \}$ is randomly chosen. The system is then left to precess for a time $t$.

(4) The position $x(t)$ is measured at the chosen time. Since the round ends here, the measurements can be destructive.

After many rounds, the average probability of finding the position of the particle to be positive is calculated as the score $P_3(T_0)$, that is,
\begin{align}\label{P3T1}
    P_3(T_0) &= \frac{1}{3} \sum_{k=0}^2 \Bqty{\Pr\bqty{ x\pqty{\tfrac{kT_0}{3}} > 0 } + \frac{1}{2}\Pr\bqty{ x\pqty{\tfrac{kT_0}{3}} = 0 }} \nonumber\\
    &= \frac{1}{3} \sum_{k=0}^2 \left\langle\pos\bqty{ x\pqty{\tfrac{kT_0}{3}}}\right\rangle,
\end{align}
where $\pos(x) = [1+\sgn(x)]/2$, with the usual convention $\sgn(x=0) = 0$.

The crucial observation is this: Under the assumption that the dynamics is a uniform precession, if the observed score satisfies $P_3(T_0) > \mathbf{P}_3^c := \frac{2}{3}$, then the system is certified to be quantum. To see why this is the case, let us consider every classical phase-space trajectory, some of which are illustrated in Fig.~\ref{fig:Harmonic_initial_states}(a). For any classical state of maximal information that is prepared in any of the orange regions at $t=0$, $x(t) > 0$ at two out of three possible times $t \in \{0,T_0/3,2T_0/3\}$ and hence will have a score of $P_3^c = \frac{2}{3}$. For those initially prepared in the blue regions, $x(t) > 0$ at one out of the three times with a score of $P_3^c = \frac{1}{3}$. Meanwhile, the fixed point $(x,p)=(0,0)$ has the score $P_3^c = \frac{1}{2}$. As a general classical state would be some convex mixture of these states of maximal information, the maximum achievable score with a classical harmonic oscillator is $\mathbf{P}^c_3 = \frac{2}{3}$.

In quantum theory, the score $P_3(T_0)$ is given by the expectation value $\ev{Q_3(T_0)}$ of the observable
\begin{equation}
    Q_3(T_0) = \frac{1}{3} \sum_{k=0}^2 \pos\bqty{ X\pqty{\tfrac{kT_0}{3}} }, \label{Q3T1}
\end{equation}
where $X(t) = e^{iHt/\hbar} X e^{-iHt/\hbar}$ is the position observable in the Heisenberg picture and $\pos(X)\ket{x} = \pos(x)\ket{x}$. Generally, the state of a quantum system is specified by a density operator $\rho$, with which $\ev{Q_3(T_0)} = \tr[\rho Q_3(T_0)]$.

There are quantum states such that $\tr[\rho Q_3(T_0)] > \mathbf{P}_3^c$. The states that give the largest quantum value are eigenvectors of $Q_3(T_0)$ for the largest eigenvalue and were discussed in \cite{tsirelson,Zaw2022}. Here, in anticipation of the energy constraints to be expounded in later sections, we consider the state that gives the maximal violation under the truncation $E_n \leq E_\mathbf{n}$ for some maximal number of excitations $\mathbf{n}$. In the original protocol, no violation is obtained for $\mathbf{n}<6$, while for $\mathbf{n}=6$ the state \begin{equation}\label{eq:harmonic-optimal-n6-state}
    \ket{\Psi_6} = \frac{1}{\sqrt{2}}\pqty{
        \frac{4}{\sqrt{21}} \ket{0} - \ket{3} + \sqrt{\frac{5}{21}}\ket{6}
    },
\end{equation}
where the $\ket{n}$ are the usual number states of the quantum harmonic oscillator, achieves the score $P_3(T_0) = \mel{\Psi_6}{Q_3(T_0)}{\Psi_6} = 0.687$. 

\begin{figure}
    \centering
    \includegraphics{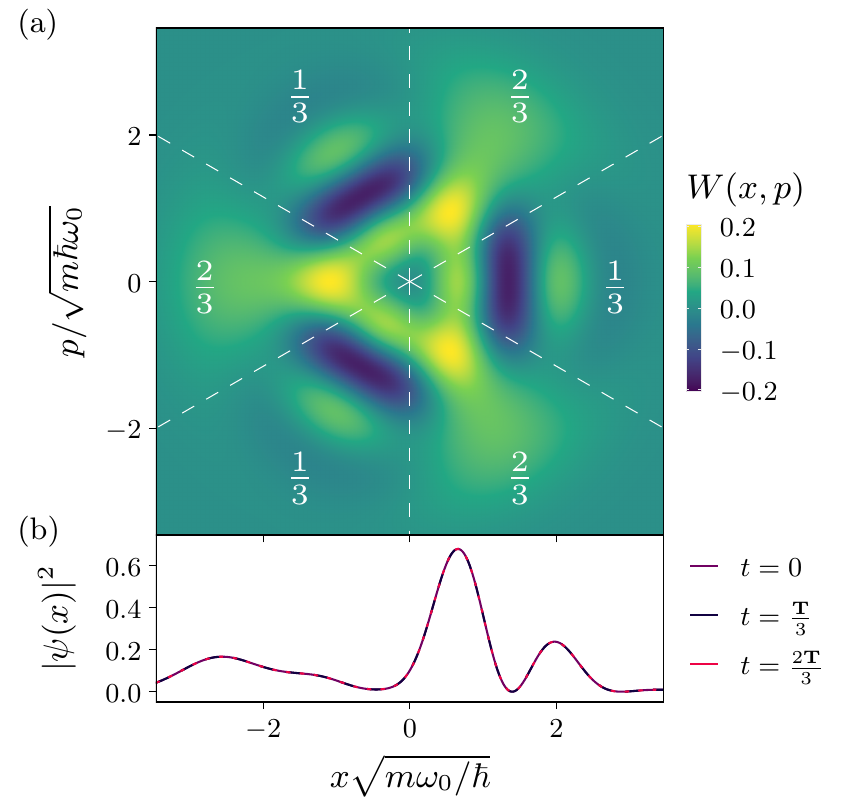}
    \caption{\label{fig:Harmonic_n6_states}For the harmonic oscillator, (a) the Wigner function and (b) the probability density at the probing times $t=0,T_0/3,2T_0/3$ of the state $\ket{\Psi_6}$ given in Eq.~\eqref{eq:harmonic-optimal-n6-state}, which violates the classical bound with the score $P_3(T_0) = 0.687 > \mathbf{P}_3^c$. The classical scores $P_3^c$ are superimposed on the corresponding regions as illustrated in Fig.~\ref{fig:Harmonic_initial_states}(a). The negativities of the Wigner function are concentrated in the regions with the lowest classical score, which augments the positivity of the Wigner function in the region with the largest classical score. This is the state that maximally violates the classical bound for the original protocol under the restriction $E_n \leq E_\mathbf{n}$ for $\mathbf{n} = 6$.}
\end{figure}

When only position measurements (or, more generally, quadrature measurements) are involved, a quantum state with a non-negative Wigner function is indistinguishable from a classical state with the same joint probability distribution. Thus, some Wigner negativity has to be present for a gap between the quantum and classical scores. In addition, this negativity must be present at the right location. The intuition is clearly conveyed by the Wigner function of $\ket{\Psi_6}$ (Fig.~\ref{fig:Harmonic_n6_states}): The negativities are concentrated in the regions with the lowest classical score ($P_3^c=\frac{1}{3}$), which augments the positivity in the regions with the highest classical score ($P_3^c=\frac{2}{3}$). Every other state that violates the classical bound also shares this general behavior.

In summary, under the assumption that its dynamics is a uniform precession, the observation $P_3(T_0) > \mathbf{P}_3^c = \frac{2}{3}$ certifies the quantumness of the system. The classical bound can be violated by preparing states that have Wigner negativities in the regions that achieve the minimum scores in the classical system.

\begin{figure}
    \includegraphics{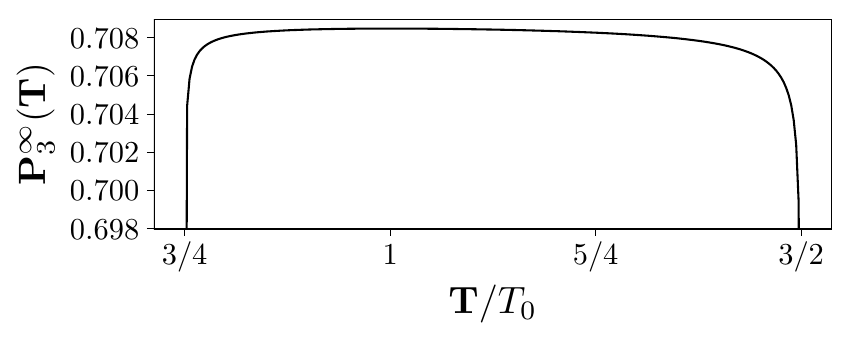}
    \caption{\label{fig:Harmonic_T}For the harmonic oscillator, the maximum quantum score $\mathbf{P}_3^\infty(\mathbf{T})$ against probing duration $\mathbf{T}$. Here $\mathbf{P}_3^\infty(\mathbf{T})$ was found by solving for the maximum eigenvalue of $Q_3(\mathbf{T})$ with the truncation $n \leq 6000$.}
\end{figure}

\subsection{Changing the probing times}\label{sec:newharm}

Now we proceed to modify step 3 of the protocol. This could be done in different ways: choosing three arbitrary times \cite{tsirelson} or probing at $K>3$ equally spaced times \cite{Zaw2022}. For the sake of this work, we focus on a single-parameter family.

(3$^\prime$) A duration $t\in\{0,\mathbf{T}/3,2\mathbf{T}/3\}$ is randomly chosen for a value $\mathbf{T}$ which may differ from the period of the precession. The system is then left to evolve for a time $t$.

The rest of the protocol is unchanged, with the score $P_3(\mathbf{T})$ now obtained by replacing $T_0$ with $\mathbf{T}$ in Eq.~\eqref{P3T1}.

Of course, the value of $\mathbf{T}$ may affect the classical bound: For example, $\mathbf{T} = 3T_0$ results in $x(0) = x(\mathbf{T}/3) = x(2\mathbf{T}/3)$ such that a state initially in the positive-$x$ plane scores $P_3(\mathbf{T}=3T_0)=1$. In that case, there cannot be any gap between the classical and quantum expectations. It is straightforward to prove that the bound $\mathbf{P}_3^c = \frac{2}{3}$ holds in the range $3T_0/4 \leq \mathbf{T} \leq 3T_0/2$ (Fig.~\ref{fig:Harmonic_initial_states}).

For the quantum observable, we analogously define $Q_3(\mathbf{T})$ by replacing $T_0$ with $\mathbf{T}$ in Eq.~\eqref{Q3T1}. As shown in Fig.~\ref{fig:Harmonic_T}, the classical bound can be violated within the range of $\mathbf{T}$ up to $\mathbf{P}^\infty_3(\mathbf{T}) := \max_\rho \tr[\rho Q_3(\mathbf{T})]$, with maximal violation for $\mathbf{T}=T_0$. At the boundary points, no violation can be found since
\begin{equation}
\begin{aligned}
Q_3(3T_0/4) &= \frac{1}{3} + \frac{1}{3}\pos[X(T_0/2)], \\
Q_3(3T_0/2) &= \frac{1}{3} + \frac{1}{3}\pos[X(0)],
\end{aligned}
\end{equation}
in which cases $\mathbf{P}^\infty_3(3T_0/4) = \mathbf{P}^\infty_3(3T_0/2) = \frac{2}{3} = \mathbf{P}_3^c$.

The flexibility in probing durations introduces a qualitative difference when we look at truncations. As mentioned before, no state with $\mathbf{n}<6$ violates the original protocol. If we maintain the truncation at $\mathbf{n}=6$, the maximum is still achieved by $\ket{\Psi_6}$ [Eq.~\eqref{eq:harmonic-optimal-n6-state}] for $\mathbf{T}=T_0$. However, by changing the probing time, we can find states with $\mathbf{n}<6$ that give a violation. Specifically, for $\mathbf{n}=4$ the state
\begin{equation}\label{eq:harmonic-optimal-n4-state}
\begin{aligned}
    \ket{\Psi_4} &=
        \sqrt{0.279}\ket{0} +
        \sqrt{0.191}\ket{1}e^{-i\theta_4} +
        \sqrt{0.121}\ket{2}e^{-i2\theta_4} \\
        &\quad {}+{}\sqrt{0.309}\ket{3}e^{-i3\theta_4} +
        \sqrt{0.100}\ket{4}e^{-i4\theta_4},
\end{aligned}
\end{equation}
where $\theta_4 = 0.215\pi$, achieves $P_3(\mathbf{T}) = \mel{\Psi_4}{Q_3(\mathbf{T})}{\Psi_4} = 0.669$ with probing duration $\mathbf{T} = 1.177T_0$. % 1613\pi/7500

\begin{figure}
    \centering
    \includegraphics{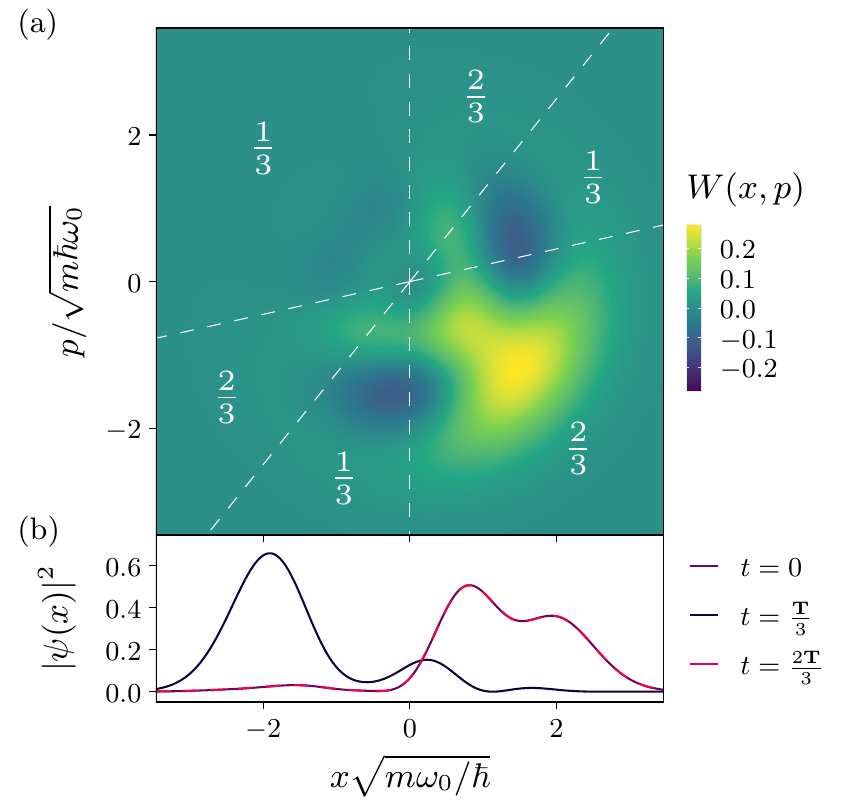}
    \caption{\label{fig:Harmonic_n4_states}For the harmonic oscillator, (a) the Wigner function and (b) the probability density at the probing times $t=0,\mathbf{T}/3,2\mathbf{T}/3$ of the state $\ket{\Psi_4}$ given in Eq.~\eqref{eq:harmonic-optimal-n4-state}, which violates the classical bound with the score $P_3(\mathbf{T}) = 0.669 > \mathbf{P}_3^c$. The probing duration is $\mathbf{T} = 1.774T_0$, where $T_0$ is the natural period of the harmonic oscillator. The classical scores $P_3^c$ are superimposed on the corresponding regions as illustrated in Fig.~\ref{fig:Harmonic_initial_states}(d). Here the negativities of the Wigner function are concentrated in the regions with the lowest classical score, which augments the positivity of the Wigner function in the region with the largest classical score. This is the state that maximally violates the classical bound for the protocol with varying probing times under the restriction $E_n \leq E_\mathbf{n}$ for $\mathbf{n} = 4$.}
\end{figure}

The Wigner function and marginal probability distribution of $\ket{\Psi_4}$ are plotted in Fig.~\ref{fig:Harmonic_n4_states}. As usual, the Wigner function is distributed in such a manner that the positive (negative) weights are concentrated in the regions where the classical score is $P_3^c = \frac{2}{3}$ ($P_3^c = \frac{1}{3}$). However, while the Wigner function of $\ket{\Psi_6}$ has the same symmetry as the protocol (the system is found in the same state at all three probing times), the positivity of $\ket{\Psi_4}$ is mostly concentrated on only one $P_3^c = \frac{2}{3}$ region. At the times $t \in \{0,2\mathbf{T}/3\}$, almost all of the probability distribution lies on the positive-$x$ axis. At time $t=\mathbf{T}/3$, most of the probability is on the negative-$x$ axis, but a small contribution from the positive-$x$ axis augments the quantum score.

The take-away message is that we can extend the notion of the probing duration $\mathbf{T}$ to something other than the natural period $T_0$ of the system, as long as we make a choice that maintains the classical bound to be $\mathbf{P}_3^c < 1$, so that a quantum-classical gap can be possible (though it might not be guaranteed). In some cases, like with the low-energy state $\ket{\Psi_4}$ with $\mathbf{n}=4$, the quantum score can be improved by using a different probing duration.

\section{Protocol for general Hamiltonians}
\label{sec:general}

Most forms of experimental trapping are approximately harmonic. In these cases, our original protocol could be straightforwardly applied to detect nonclassicality in the parameter regime where the harmonic approximation holds. This notably includes a tight upper bound on the energy \cite{harmonic-approximation}. By adopting a general approach instead of a reduction to the harmonic case, we will be able to detect nonclassicality in a much larger parameter range. In addition, we will also be able to deal with potentials that are not approximately harmonic.

\subsection{Overview of the general protocol}

We consider systems with one degree of freedom with canonical coordinates $(q,p)$ whose dynamics is generated by a time-independent Hamiltonian $H(q,p)$. The dynamics of the corresponding quantum system is given by the Weyl-quantized Hamiltonian $H(Q,P)$, where $(Q,P)$ are the canonically conjugate operators which satisfy $e^{ipQ/\hbar}e^{iqP/\hbar} = e^{-ipq/\hbar}e^{iqP/\hbar}e^{ipQ/\hbar}$, the Weyl form of the canonical commutation relation. The Weyl relation implies $\comm{Q}{P}=\comm{X}{P}=i\hbar\mathbbm{1}$ for Cartesian coordinates $(q,p)=(x,p)$ \cite{Weyl}, but is also valid with angular coordinates $(q,p)=(\phi,l)$ \cite{angular-momentum-phase-space}. For the Cartesian coordinates, this relation results in the same equations of motions of $x(t)$ and $p(t)$ for both the quantum and classical systems, regardless of the form of the Hamiltonian \cite{Moyal}.

Once again, the general protocol consists of many independent rounds. In each round, the following steps take place.

(1) The system is prepared in some state.

(2$^\prime$) After the preparation is completed, the system is decoupled from everything else and undergoes the closed dynamics specified by $H(q,p)$.

(3$^\prime$) A duration $t\in\{0,\mathbf{T}/3,2\mathbf{T}/3\}$ is randomly chosen for a specified value $\mathbf{T}$. The system is then left to evolve for a time $t$.

(4$^\prime$) The coordinate $q(t)$ is measured at the chosen time. Since the round ends here, the measurements can be destructive.

After many rounds, we calculate $P_3(\mathbf{T})$, the average probability that the coordinate is found to be positive:
\begin{equation}\label{eq:P3-definition}
\begin{aligned}
    P_3(\mathbf{T}) &:= \sum_{k=0}^{2}\Bqty{
        \Pr[q(\tfrac{k\mathbf{T}}{3})>0] + \frac{1}{2}\Pr[q(\tfrac{k\mathbf{T}}{3})=0]
    }\\
    &= \frac{1}{3} \sum_{k=0}^2 \pos\bqty{ q\pqty{\tfrac{k\mathbf{T}}{3}} }.
\end{aligned}
\end{equation}
If the measured score $P_3(\mathbf{T})$ is found to be strictly larger than the classical bound $\mathbf{P}^c_3$, then the system must be quantum. For the calculations in quantum theory, given a state $\rho$, the score is calculated by $P_3(\mathbf{T}) = \tr[\rho Q_3(\mathbf{T})]$, where
\begin{equation}\label{eq:Q3-definition}
    Q_3(\mathbf{T}) := \frac{1}{3}\sum_{k=0}^{2} \pos\bqty{Q(\tfrac{k\mathbf{T}}{3})},
\end{equation}
$\pos(Q) = \bqty{\mathbbm{1}+\sgn(Q)}/2$, and $Q(t) = e^{iHt/\hbar} Q e^{-iHt/\hbar}$.

Of course, one can more generally calculate the score with $\Pr[q(t) > q_0]$ against some reference position $q_0$ or instead consider the likelihood $\Pr[q(t) < 0]$ that the coordinate is found to be negative. In those cases, it is a simple matter to redefine $q'(t) := q(t)-q_0$ or $q'(t) := -q(t)$, respectively, which we can always do as we are free to choose our coordinates. Doing so returns it to the form given in Eq.~\eqref{eq:P3-definition}, so we will stick to this standard form.

In order to find a classical-quantum gap, we will have to impose additional constraints, which generically take the form of energy bounds $E_{\text{min}} \leq E \leq E_{\text{max}}$ (secondary assumptions may also be required for specific systems). We proceed to derive these constraints rigorously.

\subsection{Constraints in classical theory}\label{sec:classical-constraints}
\begin{figure}
    \centering
    \includegraphics{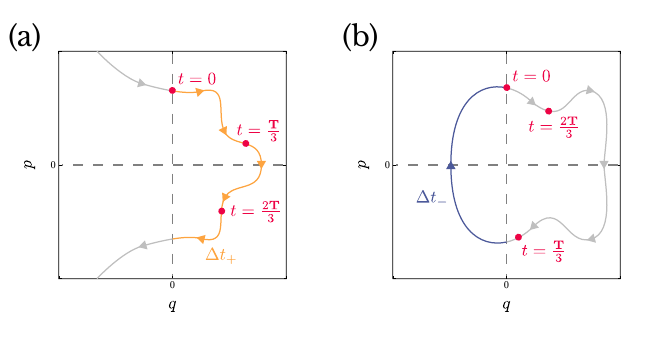}
    \caption{For the generic condition, the score $P_3=1$ occurs for certain classical trajectories (a) if all three measurements are performed before the state crosses the $p$ axis and (b) if the state crosses the $p$ axis at least twice between two measurements. Note that (a) also includes the situation where the state remains in the positive-$q$ plane for all time and (b) also includes the situation where the state crosses the $p$ axis more than twice. Also labeled are $\Delta t_+$ and $\Delta t_-$, the longest and shortest times spent in the positive- and negative-$q$ planes, respectively.}
    \label{fig:worst-classical-cases}
\end{figure}

A nontrivial classical bound is required for the protocol, which for three probing times can only be $\mathbf{P}^c_3=\frac{2}{3}$. So we need to exclude the possibility that $P_3^c=1$, i.e.,~evolutions such that $q(t)>0$ at all measured times $t \in \{0,\mathbf{T}/3,2\mathbf{T}/3\}$. Like in the case of the harmonic oscillator, $P_3^c=1$ may happen if $\mathbf{T}$ is either too short or too long; for generic dynamics, these bounds may vary with the trajectory. The situations to avoid are illustrated in Fig.~\ref{fig:worst-classical-cases}.

Formally, consider any phase-space trajectory $\gamma$, with its generalized coordinate denoted by $q_\gamma(t)$. We exclude the trivial trajectory $q_{\gamma_0}(t) =0\;\forall t$, which would lead to cumbersome notation but whose score is $P_3(\mathbf{T}) = \frac{1}{2}$ for any $\mathbf{T}$ by the definition of $P_3(\mathbf{T})$. Let us define $\Delta t_+(\gamma)$ as the longest amount of time that the trajectory $\gamma$ spends in the region where $q$ is non-negative:
\begin{widetext}
\begin{equation}\label{eq:delta-t-pos}
\Delta t_+(\gamma) := \begin{cases}
\infty & \text{if $q_\gamma(t) \geq 0 \; \forall t$}\\
\max\Bqty{
\Delta t : \exists t_0: t_0 \leq t \leq t_0+\Delta t: q_\gamma(t) \geq 0
} & \text{otherwise.}
\end{cases}
\end{equation}
Similarly, we define $\Delta t_-(\gamma)$ as the shortest amount of time that the trajectory $\gamma$ spends in the region where $q$ is strictly negative:
\begin{equation}\label{eq:delta-t-neg}
\Delta t_-(\gamma) := \begin{cases}
\infty & \text{if $q_\gamma(t) < 0\;\forall t$}\\
\min\Bqty{
    \Delta t : 
    \exists t_0: q_\gamma(t_0) \geq 0;
    q_\gamma(t_0+\Delta t) \geq 0;
    t_0 < t < t_0+\Delta t: q_\gamma(t) < 0
} & \text{otherwise.}
\end{cases}
\end{equation}
\end{widetext}
For the classical bound to be $\mathbf{P}^c_3=\frac{2}{3}$, we should restrict ourselves to a set $\Gamma$ of classical trajectories and a probing time such that
\begin{equation}\label{eq:general-classical-condition}
\begin{gathered}
\max_{\gamma \in \Gamma \setminus \{\gamma_0\}} \frac{3}{2} \Delta t_+(\gamma) \leq \mathbf{T} \leq \min_{\gamma \in \Gamma \setminus \{\gamma_0\}} 3\Delta t_-(\gamma)\,.
\end{gathered}
\end{equation}
As a consistency check for the harmonic oscillator, $\Delta t_+(\gamma)=\Delta t_-(\gamma)=T_0/2$ for all $\gamma$, recovering the previous condition on $\mathbf{T}$.

Quantitatively, Eq.~\eqref{eq:general-classical-condition} provides the exact inequality for which a nontrivial classical bound is achieved. Qualitatively, it expounds two conditions. First, the lower bound states that the particle cannot spend an arbitrarily long time in the positive region, which means that there must be some barrier on the positive side of the origin. Second, the upper bound states that the particle cannot spend an arbitrarily short time in the negative region. Specifically, the barrier on the negative side of the origin cannot be too strong, lest the particle be pushed back to the positive side too rapidly even in the classical system, which reduces the quantum-classical gap. Ultimately, to obtain nontrivial classical bounds, some form of trapping is required on the positive axis, without the trapping being too strong on the negative axis. We stress once again that this trapping need not be approximately harmonic; the only requirement is Eq.~\eqref{eq:general-classical-condition}.

\subsection{Energy constraints}
For use in our protocol, the constraint~\eqref{eq:general-classical-condition} on trajectories needs to be translated into one or several constraints on quantities that are observables in both quantum and classical theory. The obvious choice is energy. Indeed, when the Hamiltonian is time independent, the energy is an integral of motion; hence $H(q_\gamma(t),p_\gamma(t)) = E$ along any trajectory $\gamma$. For systems with one degree of freedom, the equation $H(q,p) = E$ defines a one-dimensional curve. If $E$ is not degenerate, it uniquely specifies a phase-space trajectory. If $E$ is degenerate, as phase-space trajectories cannot cross, $H(q,p) = E$ specifies a set $\{\gamma_k\}$ of distinct trajectories.

Now let us define
\begin{equation}\label{eq:energy-extremal-period}
\begin{aligned}
\Delta t_+(E) &:= \max_{\gamma\in\Gamma \setminus \{\gamma_0\}:H(q_\gamma,p_\gamma)=E}\Delta t_+(\gamma) \\
\Delta t_-(E) &:= \min_{\gamma\in\Gamma \setminus \{\gamma_0\}:H(q_\gamma,p_\gamma)=E}\Delta t_-(\gamma).
\end{aligned}
\end{equation}
Like previously, $\Gamma$ is the set of all classical trajectories under consideration. Without any additional assumptions, this is the set of all possible trajectories, while secondary assumptions would place restrictions on $\Gamma$. 

Using Eq.~\eqref{eq:energy-extremal-period}, Eq.~\eqref{eq:general-classical-condition} can be rewritten in terms of the energy bounds $E_{\text{min}} \leq E \leq E_{\text{max}}$,
\begin{equation}\label{eq:energy-classical-condition}
    \max_{E_{\text{min}}\leq E \leq E_{\text{max}}} \frac{3}{2} \Delta t_+(E) \leq \mathbf{T} \leq \min_{E_{\text{min}}\leq E \leq E_{\text{max}}} 3\Delta t_-(E).
\end{equation}
Therefore, $E_{\text{min}}$ and $E_{\text{max}}$ should be chosen such that Eq.~\eqref{eq:energy-classical-condition} is satisfied. In the quantum system, this condition manifests as a projection of the full Hilbert space onto a smaller, oftentimes finite, subspace spanned by the energy eigenstates whose energies lie within that range.

\subsection{Summary of procedure}\label{sec:summary-procedure}
Let us summarize the procedure for finding the probing duration $\mathbf{T}$ with respect to some energy bounds $E_{\text{min}}$ and $E_{\text{max}}$.

(1) Identify $\Gamma$, the set of classical trajectories under consideration. Secondary assumptions might restrict $\Gamma$ to a subset of all possible trajectories.

(2)Find $\Delta t_\pm (E)$ as a function of energy.

(3) Invert Eq.~\eqref{eq:energy-classical-condition} to obtain inequalities for $E_{\text{min}}$ and $E_{\text{max}}$ as functions of $\mathbf{T}$. Choose a valid $\mathbf{T}$ such that $E_{\text{min}}\leq E_{\text{max}}$.

(4) Construct $Q_3(\mathbf{T})$ in the energy-eigenstate basis, restricted to the subspace $\{\ket{E_n}\}$, where $E_{\text{min}} \leq E_n \leq E_{\text{max}}$.

(5) Solve for the maximum eigenvalue of $Q_3(\mathbf{T})$.

If the maximum eigenvalue $\mathbf{P}_3^\infty$ is larger than $\mathbf{P}_3^c = \frac{2}{3}$, then there are states that would allow us to certify the quantumness of the system by performing the protocol with probing duration $\mathbf{T}$. 

Let us now explain how one may go about steps 2 and 4 in some generality.

\subsubsection{Simplifications for Step 2}\label{sec:simplify-probing-period}
Depending on symmetries present in the Hamiltonian, some simplifications can be made when calculating $\Delta t_\pm$.

\begin{figure}
    \centering
    \includegraphics{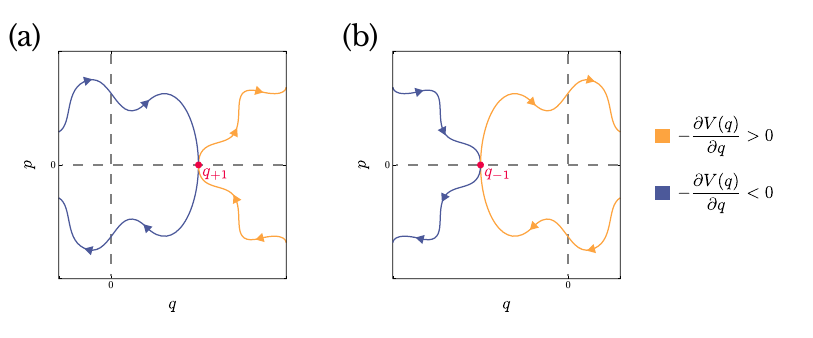}
    \caption{\label{fig:General_NonrelativisticBounds}For the general approach, possible trajectories for nonrelativistic Hamiltonians when the only solutions to $E-V(q)=0$, excluding the fixed point $(q=0,p=0)$, are $q_{-1}$ or $q_{+1}$, or both. (a) If $q_{-1}$ is not a solution, there are two possibilities: if $\frac{dp}{dt}\rvert_{q_{+1},p=0} = -\frac{\partial V(q)}{\partial q}\rvert_{q_{+1}} \geq 0$, the trajectory will never enter the negative-$q$ plane and hence $\Delta t_-(E) = 0$; if $-\frac{\partial V(q)}{\partial q}\rvert_{q_{+1}} < 0$, the trajectory enters the negative-$q$ plane, but as it does not intersect the $q$ axis again, it remains in the negative $q$ plane and hence $\Delta t_-(E) = 0$. (b) If $q_{-1}$ is a solution, there are also two possibilities: If $-\frac{\partial V(q)}{q}\rvert_{q_{-1}} > 0$, the trajectory will intersect the $p$ axis at the points $p_-$ and $p_+$ with $p_- < p_+$, and hence $\Delta t_-(E)$ can be found as an integral over the path from $(q=0,p_-)$ to $(q=0,p_+)$, as given in Eq.~\eqref{eq:period-numerical-neg}. If $-\frac{\partial V(q)}{q}\rvert_{q_{-1}} \leq 0$, the trajectory will never enter the positive-$q$ plane, and hence $\Delta t_-(E) = \infty$. Analogous expressions for $\Delta t_+(E)$ can be found by a reflection $q \to -q$.}
\end{figure}

\textit{(1) $H(q,p)$ even in $q$}. When $H(-q,p) = H(q,p)$, every collection of trajectories $\{\gamma_k\}$ with energy $E$ is symmetric about the $p$ axis. Therefore, the amounts of time spent in the positive- and negative-$q$ planes are the same, so $\Delta t_+(E) = \Delta t_-(E)$. If $\gamma$ is a closed path that intersects the $p$ axis and has period $T_0(\gamma)$, then $\Delta t_\pm(\gamma) = T_0(\gamma)/2$.

\textit{(2) $H(q,p)$ even in $p$}. When $H(q,-p) = H(q,p)$, every trajectory $\gamma$ is symmetric about the $q$ axis. Hence, it is adequate to consider the trajectory $(q_\gamma(t),p_\gamma(t))$ that starts from $p(0)=0$ for the times $0 \leq t < \infty$, with the solution for $t<0$ obtained with a reflection. Hence, given the initial conditions $p(0)=0$ and $x(0) < 0$ [$p(0)=0$ and $x(0) > 0$], $\Delta t_-(\gamma)$ [$\Delta t_+(\gamma)$] is twice the time taken for the trajectory to reach $x = 0$.

\emph{(3) Hamiltonian is nonrelativistic}. If $H(q,p) = p^2/2\mu + V(q)$, the solutions to $E-V(q) = 0$ are the points where a trajectory with energy $E$ intersects the $q$ axis, which we will denote by $\cdots < q_{-2} < q_{-1} < 0 < q_{+1} < q_{+2} < \cdots$. Note that all possible trajectories with energy $E$ can only intersect the $p$ axis at the two points $p_\pm = \pm\sqrt{2\mu[E-V(0)]}$. Furthermore, since $H(q,p)$ is even in $p$, the fact that every trajectory is symmetric about the $q$ axis implies that the trajectory that contains the points $(q,0)$ and $(0,p_+)$ must also contain the point $(0,p_-)$. Therefore, if $q_{+2}$ is a solution, then the two trajectories that intersect the $q$ axis at $q_{+1}$ and $q_{+2}$ cannot both cross the $p$ axis. In other words, at least one of the two trajectories must remain indefinitely in the positive-$q$ plane such that $\Delta t_{+}(E) = \infty$. If $q_{-2}$ is a solution, an analogous reasoning gives $\Delta t_{-}(E) = \infty$.

Meanwhile, if $q=0$ is a solution to $E-V(q)=0$, there are three possibilities: $(q=0,p=0)$ is a fixed point, $-\frac{\partial V(q)}{\partial q}\rvert_{q=0} > 0$, or $-\frac{\partial V(q)}{\partial q}\rvert_{q=0} < 0$. If $(q=0,p=0)$ is a fixed point, it is excluded from the definition of $\Delta t_{\pm}(E)$ and can be ignored. If $-\frac{\partial V(q)}{\partial q}\rvert_{q=0} = \frac{dp}{dt}\rvert_{(q=0,p=0)} > 0$, the trajectory remains in the positive-$q$ plane, and hence $\Delta t_{+}(E) = \infty$. Conversely, $-\frac{\partial V(q)}{\partial q}\rvert_{q=0} = \frac{dp}{dt}\rvert_{(q=0,p=0)} < 0 \implies \Delta t_{-}(E) = \infty$.

These arguments imply that the only situation where $\Delta t_{\pm}(E)$ is finite is when, excluding the fixed point $(q=0,p=0)$, either $q_{-1}$ or $q_{+1}$ or both $q_{-1}$ and $q_{+1}$ are the only solutions of $E-V(q)=0$. If that is the case, the possible trajectories are illustrated in Fig.~\ref{fig:General_NonrelativisticBounds}, which gives rise to the following formulas for $\Delta t_\pm(E)$, which allows us to calculate $\Delta t_{\pm}(E)$ with numerical integration even when the exact trajectories cannot be solved:
\begin{widetext}
\begin{align}\label{eq:period-numerical-neg}
\Delta t_-(E) &= \begin{cases}
0 & \text{if $q_{-1}$ is not a solution and $\left.-\frac{\partial V(q)}{\partial q}\right\rvert_{q_{+1}} \geq 0$} \\
\sqrt{\frac{\mu}{2}} \int_{q_{-1}}^0 \frac{\dd{q}}{\sqrt{E-V(q)}} & \text{if $q_{-1}$ is a solution and $\left.-\frac{\partial V(q)}{\partial q}\right\rvert_{q_{-1}} > 0$} \\
\infty & \text{otherwise,}
\end{cases} \\\label{eq:period-numerical-pos}
\Delta t_+(E) &= \begin{cases}
0 & \text{if $q_{+1}$ is not a solution and $-\left.\frac{\partial V(q)}{\partial q}\right\rvert_{q_{-1}} \leq 0$} \\
\sqrt{\frac{\mu}{2}} \int_{0}^{q_{+1}}\dd{q} \frac{1}{\sqrt{E-V(q)}} & \text{if $q_{+1}$ is a solution and $-\left.\frac{\partial V(q)}{\partial q}\right\rvert_{q_{+1}} < 0$} \\
\infty & \text{otherwise.}
\end{cases}
\end{align}

\subsubsection{Simplifications for Step 4}\label{sec:simplify-quantum-observable}
For the eigenstates $\ket{E_n}$ of $H$ that belong to the discrete part of the energy spectrum, we need to compute the matrix elements
\begin{equation}
\begin{aligned}
    \mel{E_n}{Q_3}{E_{n'}} = \frac{1}{2}\delta_{n,n'} + \frac{1}{6}\sum_{k=0}^2 e^{i k\mathbf{T}\theta_{n,n'}} \mel{E_n}{\sgn(Q)}{E_{n'}},
\end{aligned}
\end{equation}
where $\theta_{n,n'}:=(E_n-E_{n'})/3\hbar$. In the examples studied in this paper, the spectrum of $Q$ is continuous with support $q \in [q_{\text{min}},q_{\text{max}}]$, where $q_{\text{min}} < 0 < q_{\text{max}}$. The matrix elements of $\sgn(Q)$ are found with the integration
\begin{equation}
\mel{E_n}{\sgn(Q)}{E_{n'}} = \int_{0}^{q_{\text{max}}}\dd{q} \braket{E_n}{q}\!\!\braket{q}{E_{n'}} - \int_{q_{\text{min}}}^0\dd{q} \braket{E_n}{q}\!\!\braket{q}{E_{n'}}.
\end{equation}
Depending on symmetries present in the Hamiltonian, some simplifications can be made when calculating the matrix elements of $\sgn(Q)$. Notably, if $H(Q,P)$ commutes with the parity operator $\Pi$, where $\Pi Q \Pi = - Q$, then there exists a set of eigenvectors $\ket{E_n}$ of $H$ that are also eigenvectors of $\Pi$. For those states, $\ev{\sgn(-Q)}{E_n}=\ev{\Pi\sgn(Q)\Pi}{E_n} = (\pm 1)^2\ev{\sgn(Q)}{E_n}$, that is, $\ev{\sgn(Q)}{E_n} = 0$.

\textit{Matrix elements for nondegenerate energies of a nonrelativistic Hamiltonian}. This simplification for systems with a nonrelativistic Hamiltonian $H(Q,P) = P^2/2\mu + V(Q)$ is due to \citet{Wronskian-trick}: Given that $Q$ has the continuous spectrum stated at the start of this section and its eigenstates $\ket{q}$ are acted upon by $P$ as $\bra{q}P = -i\hbar \frac{\partial}{\partial q}\bra{q}$, then, for $E_n \neq E_{n'}$,
\begin{align}
    \mel{E_n}{\sgn(Q)}{E_{n'}} {}={} &\frac{\hbar^2}{2\mu(E_n-E_{n'})}\Bigg(
        \braket{E_n}{q_{\text{max}}}\left.\frac{\partial\braket{q}{E_{n'}}}{\partial q}\right\rvert_{q=q_{\text{max}}}
        + \braket{E_n}{q_{\text{min}}}\left.\frac{\partial\braket{q}{E_{n'}}}{\partial q}\right\rvert_{q=q_{\text{min}}} \nonumber\\\label{eq:Wronskian-trick}
        &{}-{}
        \braket{q_{\text{max}}}{E_{n'}}\left.\frac{\partial\braket{E_n}{q}}{\partial q}\right\rvert_{q=q_{\text{max}}} -
        \braket{q_{\text{min}}}{E_{n'}}\left.\frac{\partial\braket{E_n}{q}}{\partial q}\right\rvert_{q=q_{\text{min}}} \\
        &{}-{}2\braket{E_n}{q=0}\left.\frac{\partial\braket{q}{E_{n'}}}{\partial q}\right\rvert_{q=0} + 2\braket{q=0}{E_{n'}}\left.\frac{\partial\braket{E_n}{q}}{\partial q}\right\rvert_{q=0}\nonumber
    \Bigg).
\end{align}
\end{widetext}
With Eq.~\eqref{eq:Wronskian-trick}, the matrix elements of $Q_3(\mathbf{T})$ for $E_n \neq E_{n'}$ can be found with just the values of the wave functions $\braket{q}{E_n}$ and $\braket{q}{E_{n'}}$, as well as their first derivatives, at the points $q \in \{q_{\text{min}},0,q_{\text{max}}\}$.

\section{Four case studies}

In the remainder of the paper, we study four specific examples. The first three examples, i.e., Kerr nonlinear systems (Sec.~\ref{sec:example-kerr}), the pendulum (Sec.~\ref{sec:example-pendulum}), and the Morse potential (Sec.~\ref{sec:example-morse}), are all approximately harmonic at low anharmonicities. The Morse potential admits unbound trajectories; yet we show that the protocol can still be performed with that system. For the last example, we study the particle in an infinite well (Sec.~\ref{sec:example-infinite-well}). This toy model serves to demonstrate the applicability of the protocol to general bound systems, even when the system is not approximately harmonic.

\subsection{Kerr-nonlinear systems}\label{sec:example-kerr}

As a first example, we consider a perturbation of the harmonic oscillator. We choose a form that describes both nonlinear optical systems \cite{optical-kerr-effect} and superconducting transmon systems \cite{transmon} up to the first nonlinear order: \begin{align}
    &H(Q=X,P=P) \nonumber\\\label{eq:kerr-quantum-hamiltonian}
    &\quad{}={} \hbar\omega_0\Bqty{\bqty{1 + \frac{\alpha}{2}\pqty{a^\dag a + \frac{1}{2}}}\pqty{a^\dag a + \frac{1}{2}} + \frac{3}{8}\mathbbm{1}} \\
    &\quad{}={} \hbar\omega_0(1 + \alpha) \pqty{a^{\dag} a + \frac{1}{2}\mathbbm{1}} + \hbar\omega_0 \frac{\alpha}{2} a^{\dag 2} a^2,\nonumber
\end{align}
where $a = \sqrt{{m\omega_0}/{2\hbar}}X + iP/{\sqrt{2m\hbar\omega_0}}$ is the annihilation operator defined in the usual way. The corresponding classical Hamiltonian is
\begin{equation}\label{eq:kerr-classical-hamiltonian}
    H(q=x,p=p) = H_0(x,p) + \frac{\alpha}{2\hbar\omega_0}\bqty{H_0(x,p)}^2,
\end{equation}
where $H_0(x,p) = {p^2}/{2m} + {m\omega_0^2 x^2}/{2}$ is the Hamiltonian of a harmonic oscillator. We obtain a nonstandard classical Hamiltonian, with more-than-quadratic powers of $p$, as we have used Eq.~\eqref{eq:kerr-quantum-hamiltonian} as the reference dynamics.

Concerning the anharmonicity parameter $\alpha$, negative anharmonicities are physically unfeasible, as the energy of the system would be unbounded from below. However, noting that $\alpha < 0$ sometimes appear as low-order approximations of other Hamiltonians when $H_0(x,p)/\hbar\omega_0 \ll 1$, we will allow negative values of $\alpha$ with the secondary assumption that only trajectories with $H_0(x,p)/\hbar\omega_0 \leq 1/\abs{\alpha}$ are permitted.

\subsubsection{Classical bounds}
For the trajectories under consideration, the Hamiltonian in Eq.~\eqref{eq:kerr-classical-hamiltonian} permits energies within the range
\begin{equation}
    0 \leq E/\hbar\omega_0 \leq \begin{cases}
    \infty & \text{for $\alpha \geq 0$} \\
    1/2\abs{\alpha} & \text{otherwise.}
    \end{cases}
\end{equation}
It can be easily verified that the solution to the classical equations of motion for a state with energy $E$ is
\begin{equation}\label{eq:kerr-classical-solutions}
\begin{gathered}
    x(t) = x(0)\cos[\omega(E) t] + \frac{p(0)}{m\omega}\sin[\omega(E)t],\\
    \text{where}\;\omega(E) = \sqrt{1+\frac{2\alpha E}{\hbar\omega_0}}\omega_0.
\end{gathered}
\end{equation}
Each phase-space trajectory is harmonic, uniquely identified by the energy $E$, and has an energy-dependent period $2\pi/\omega(E)$. In addition, since the Hamiltonian is even in $x$, using Sec.~\ref{sec:simplify-probing-period},
\begin{equation}
\Delta t_+(E) = \Delta t_-(E) =  \frac{2\pi/\omega_0}{2\sqrt{1+{2\alpha E}/{\hbar\omega_0}}}.
\end{equation}
Define $\mathbf{T} =: \tau \frac{2\pi}{\omega_0}$, with $\tau$ the ratio between $\mathbf{T}$ and the natural period of the unperturbed harmonic oscillator. Then the possible values of $E_{\text{min}}$ and $E_{\text{max}}$ that satisfy the condition in Eq.~\eqref{eq:energy-classical-condition} for the given $\tau$ are
\begin{equation}\label{eq:kerr-conditions}
\begin{aligned}
    \frac{E_{\text{min}}}{\hbar\omega_0} &\geq \begin{cases}
        \frac{1}{2\alpha}\pqty{\frac{9}{16\tau^2} - 1} & \text{if $\alpha \geq 0$} \\
        \frac{1}{2\alpha}\pqty{\frac{9}{4\tau^2} - 1} & \text{if $\alpha < 0$,}
    \end{cases} \\
    \frac{E_{\text{max}}}{\hbar\omega_0} &\leq \begin{cases}
        \frac{1}{2\alpha}\pqty{\frac{9}{4\tau^2} - 1} & \text{if $\alpha \geq 0$} \\
        \frac{1}{2\alpha}\pqty{\frac{9}{16\tau^2} - 1} & \text{if $\alpha < 0$.}
    \end{cases}
\end{aligned}
\end{equation}
In both cases, choosing the equality maximizes the number of classical trajectories that satisfy $P_3^c \leq \mathbf{P}_3^c$. As $E \geq 0$ for all classical trajectories under consideration, we cannot allow $E_{\text{max}}$ to be negative, as that would exclude all possible states. In order to also include all classical states with $0 \leq E \leq E_{\text{max}}$, we finally arrive at the restriction $\frac{3}{4} \leq \tau \leq \frac{3}{2}$.

\subsubsection{Quantized system}
From Eq.~\eqref{eq:kerr-quantum-hamiltonian} it is clear that the Kerr nonlinear Hamiltonian shares the same energy eigenstates $\ket{n}$ as the harmonic oscillator, with the eigenvalue
\begin{equation}
    H(X,P)\ket{n} = \underbrace{\hbar\omega_0\Bqty{\bqty{1 + \frac{\alpha}{2}\pqty{n + \frac{1}{2}}}\pqty{n + \frac{1}{2}} + \frac{3}{8}}}_{=:E_n}\ket{n}.
\end{equation}
\begin{widetext}
So the matrix elements of $Q_3$ as defined in Eq.~\eqref{eq:Q3-definition} for the Kerr-nonlinear Hamiltonian are
\begin{equation}
\mel{n}{Q_3}{n'}
= \begin{cases}
\displaystyle\frac{(-1)^{({n'-n-1})/{2}} 2^{-[({n'+n})/{2}-1]}
\pqty{1 + e^{i\tau\theta_{n,n'}} + e^{i2\tau\theta_{n,n'}}}
}{6(n'-n)}
\sqrt{
    \frac{n'}{\pi}
    \pmqty{
        n\\
        \frac{n}{2}
    }
    \pmqty{
        n'-1\\
        \frac{n'-1}{2}
    }
} & \text{for $n$ even, $n'$ odd}\\
\text{as above with $n\leftrightarrow n'$} & \text{for $n'$ even, $n$ odd}\\[1.5ex]
\displaystyle\frac{1}{2}\delta_{n,n'} & \text{otherwise,}
\end{cases}
\end{equation}
where $\theta_{n,n'} = 2\pi(n-n')\bqty{1 + \frac{\alpha}{2}\pqty{n +n' + 1}}/3$ and we have used the matrix elements of $\mel{n}{\sgn(X)}{n'}$ from \cite{Zaw2022}.
\end{widetext}
Meanwhile, the conditions in Eq.~\eqref{eq:kerr-conditions} translate to the truncation of the Hilbert space spanned by $\{\ket{n}\}_{n=0}^\infty$ to
\begin{subequations}\label{eq:truncate-hilbert-space}
\begin{align}
n &\geq
\begin{cases}
\sqrt{\pqty{\frac{3}{4\alpha\tau}}^2-\frac{3}{4}} - \frac{1}{\alpha} - \frac{1}{2} & \text{if $\alpha \geq 0$} \\
-\sqrt{\pqty{\frac{3}{2\alpha\tau}}^2-\frac{3}{4}} - \frac{1}{\alpha} - \frac{1}{2} & \text{otherwise},
\end{cases} \\
n &\leq \begin{cases}
\sqrt{\pqty{\frac{3}{2\alpha\tau}}^2-\frac{3}{4}} - \frac{1}{\alpha} - \frac{1}{2}
& \text{if $\alpha \geq 0$}\\
-\sqrt{\pqty{\frac{3}{4\alpha\tau}}^2-\frac{3}{4}} - \frac{1}{\alpha} - \frac{1}{2}
& \text{otherwise}.
\end{cases}
\end{align}
\end{subequations}
As such, we can construct $Q_3$ explicitly with the truncation~\eqref{eq:truncate-hilbert-space} and find the maximum quantum score $\mathbf{P}_3^\infty$ by solving for the largest eigenvalue of $Q_3$.

\subsubsection{Quantum violation}
\begin{figure}
    \centering
    \includegraphics{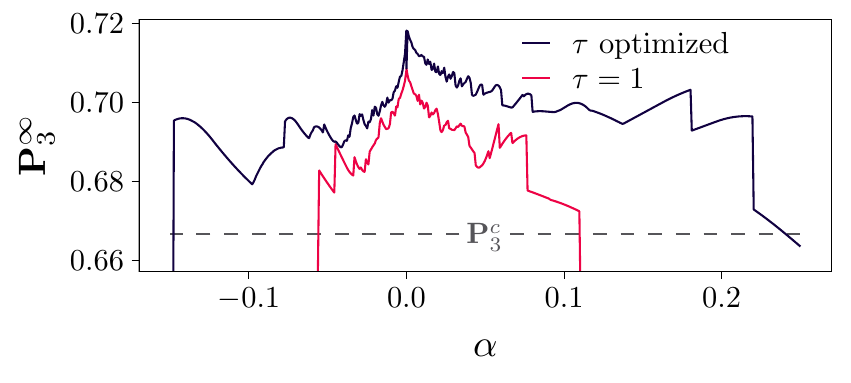}
    \caption{\label{fig:Kerr_full_range}For the Kerr nonlinear system, the maximum quantum score against anharmonicity, for $\tau=1$ and the score maximized over $\frac{3}{4}\leq\tau\leq\frac{3}{2}$. A larger violation is possible when the system is weakly anharmonic, but this advantage is eventually offset due to the energy constraints set by Eq.~\eqref{eq:truncate-hilbert-space}. The truncation in that equation also explains the discontinuous jumps.}
\end{figure}
The maximum quantum score is plotted against anharmonicity in Fig.~\ref{fig:Kerr_full_range}. For each value of $\alpha$, we plot $\mathbf{P}_3^\infty$ both for $\tau=1$ and for the optimal choice of probing duration $\frac{3}{4} \leq \tau \leq \frac{3}{2}$ (the two values coincide at $\alpha=0$, as they should). The discontinuous jumps are due to the truncation of the Hilbert space, which changes discretely according to Eq.~\eqref{eq:truncate-hilbert-space}.

As a general trend of the relationship between the quantum score and anharmonicity, we find that $\mathbf{P}_3^\infty$ is maximal around $\alpha \sim 0$ and decreases as $\abs{\alpha}$ increases. The reason for the extremal behavior is clear: As $\abs{\alpha}$ becomes too large, the truncation due to Eq.~\eqref{eq:truncate-hilbert-space} reduces the Hilbert space dimension to an extent where there can no longer be any violation.

In short, when the magnitude of the anharmonicity becomes too large, the trapping becomes either too strong or too weak: For the reasons discussed qualitatively in Sec.~\ref{sec:classical-constraints}, there is no longer be a gap between the quantum and classical scores.

\subsubsection{Comparison with the harmonic approximation}\label{sec:kerr-weak-regime}
\begin{figure}
    \centering
    \includegraphics[width=\columnwidth]{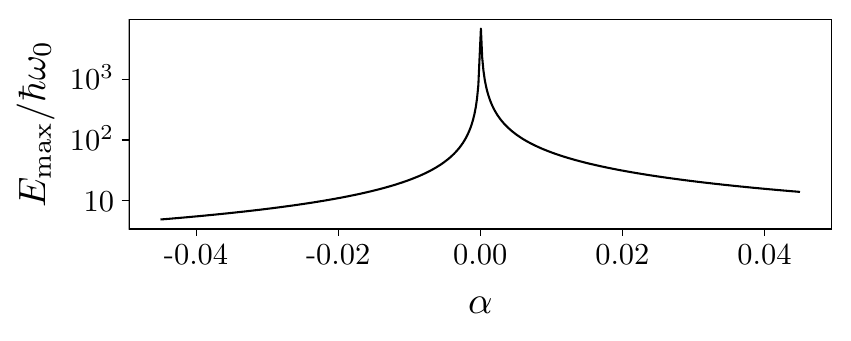}
    \caption{For the Kerr nonlinear system, the maximum energy bound against anharmonicity, for $\tau=1$. To prepare states with $E/\hbar\omega_0 \sim 1$, the requirement $E_{\text{max}} \geq 10$ is met by anharmonicities $\abs{\alpha} \leq 0.02$. There is an asymptotic behavior $E_{\text{max}}/\hbar\omega_0\to\infty$ as $\alpha\to0$.}
    \label{fig:Kerr_E_range}
\end{figure}

In this section we compare the quantum scores of a weakly anharmonic system with and without the harmonic approximation. First we need to quantify ``weak.'' To do this, we note the following. On the one hand, in order to have a violation we need to consider states with at least a few excitations [recall states $\ket{\Psi_6}$ of Eq.~\eqref{eq:harmonic-optimal-n6-state} or states $\ket{\Psi_4}$ of Eq.~\eqref{eq:harmonic-optimal-n4-state}]; so we are looking at states with $E\sim \hbar\omega_0$. On the other hand, these states should be of low energy in the potential, i.e.,~$E\ll E_{\text{max}}$. By setting $E_{\text{max}}/\hbar\omega_0 \geq 10$ we find that weak anharmonicity in our context means $\abs{\alpha} \leq 0.02$ (Fig.~\ref{fig:Kerr_E_range}).

Now, for fixed truncation of excitation $\mathbf{n}=6$, we study the role of $\alpha$ in three scenarios.

(i) Find the state $\ket{\Psi}$ and time $\tau$ that violates the classical bound maximally. Here we take the anharmonicity of the system fully into account.

(ii) Fix the state as $\ket{\Psi_6}$ from the harmonic approximation [Eq.~\eqref{eq:harmonic-optimal-n6-state}], but look for the probing time $\tau$ that maximizes the violation under the anharmonic dynamics.

(iii) Fix the state as $\ket{\Psi_6}$ and set $\tau=1$, that is, import naively the state and parameter from the harmonic approximation.

\begin{figure}
    \includegraphics{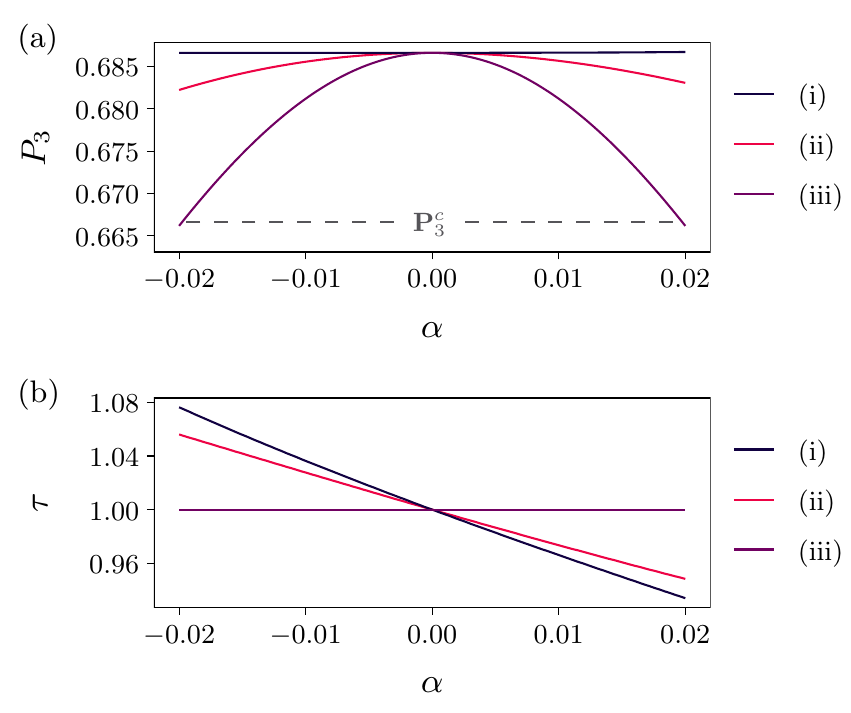}
    \caption{For the Kerr nonlinear system. (a) Score $P_3$ as a function of the anharmonicity $\alpha$ for low-energy states with $n \leq 6$. Scenario (i) corresponds to the optimal state $\ket{\Psi}$ and time $\tau$ with the largest violation, (ii) corresponds to using $\ket{\Psi_6}$ from Eq.~\eqref{eq:harmonic-optimal-n6-state} with the optimal $\tau$, and (iii) uses $\ket{\Psi_6}$ and $\tau=1$, treating the anharmonic system as if it were a harmonic one. (b) Optimal $\tau$ used for each scenario for every value of $\alpha$.}
    \label{fig:Kerr_n6}
\end{figure}

\begin{figure}
    \includegraphics{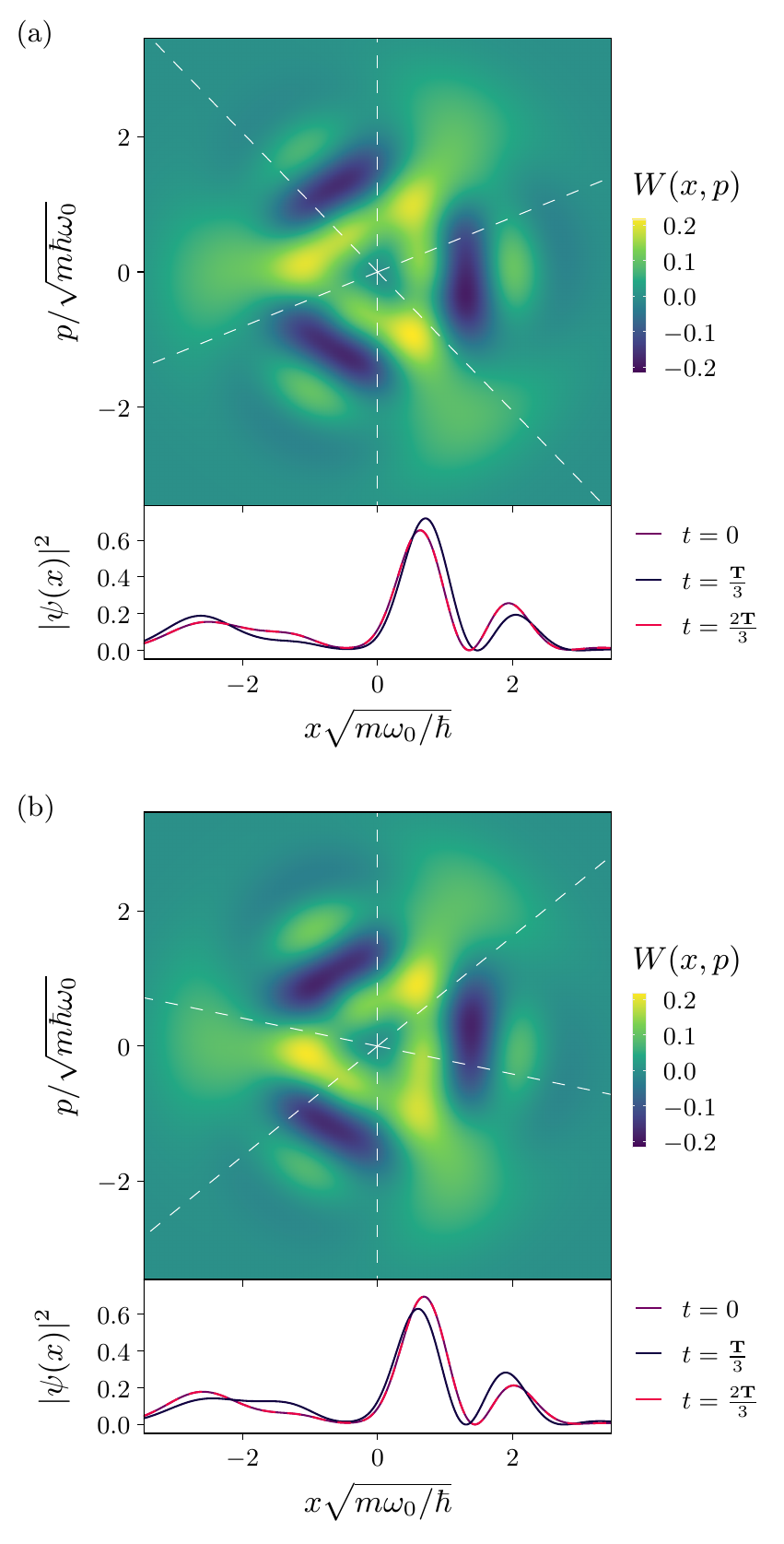}
    \caption{\label{fig:Kerr_n6_states}For the Kerr nonlinear system, the Wigner function and marginal probability distribution at times $t \in \{0,\mathbf{T}/3,2\mathbf{T}/3\}$ of the $\mathbf{n} = 6$ optimal state with (a) $\alpha=0.02$ and (b) $\alpha = -0.02$. The optimal states in the weakly anharmonic cases are slightly distorted versions of the optimal state with $\mathbf{n} = 6$ in the harmonic case (plotted in Fig.~\ref{fig:Harmonic_n6_states}). The classical score regions superimposed on the Wigner function are those that correspond to the probing duration $\tau$ (see Fig.~\ref{fig:Harmonic_initial_states}).}
\end{figure}

\begin{figure}
    \includegraphics{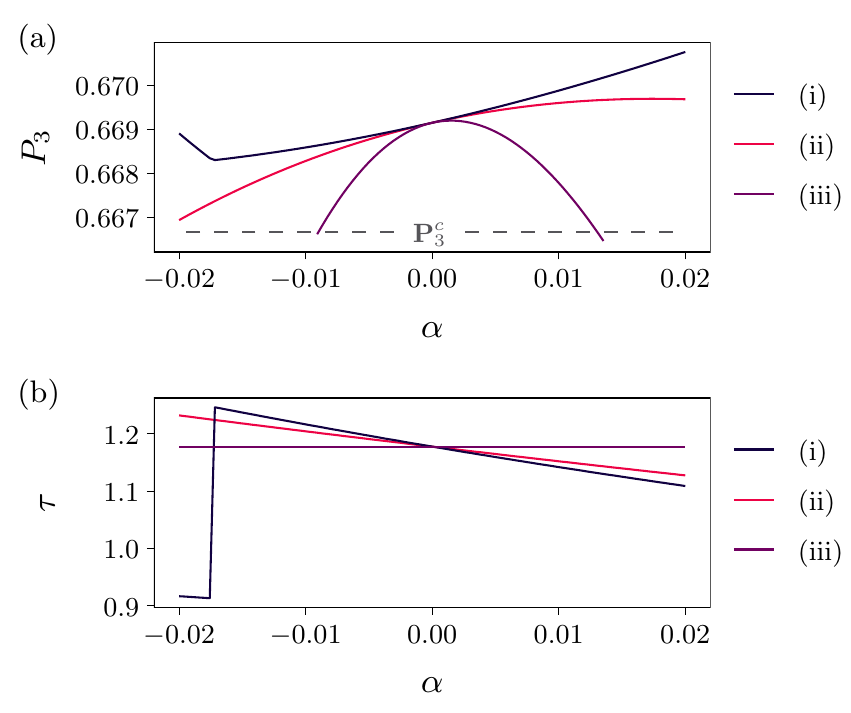}\hspace{2em}
    \caption{Kerr nonlinear system. (a) Score $P_3$ as a function of the anharmonicity $\alpha$ for the state with truncation $n \leq 4$. Scenario (i) corresponds to the optimal state $\ket{\Psi}$ and time $\tau$ with the largest violation, (ii) corresponds to using $\ket{\Psi_4}$ from Eq.~\eqref{eq:harmonic-optimal-n4-state} with the optimal $\tau$, and (iii) uses $\ket{\Psi_4}$ and $\tau=1.774$, treating the anharmonic system as if it were a harmonic one. (b) Optimal $\tau$ for each value of $\alpha$.}
    \label{fig:Kerr_n4}
\end{figure}

The scores and probing durations for the three scenarios are plotted in Fig.~\ref{fig:Kerr_n6}. We find that the quantum score decreases with the magnitude of the anharmonicity, albeit the decrease is negligible if everything is optimized for the dynamics [scenario (i)]. The optimal probing duration decreases with $\alpha$ increasing: Since the period of a fixed-energy state increases with $\alpha$, decreasing the probing duration offsets this increase. For the Wigner function of the optimal state for scenario (i) at different anharmonicities in Fig.~\ref{fig:Kerr_n6_states}, we observe that the states at $\alpha\neq0$ are distorted forms of the state $\ket{\Psi_6}$ at $\alpha=0$, with barely perceptible differences. In addition, comparing scenarios (ii) and (iii), we see that the quantum violation is greatly increased by optimizing $\tau$, even if the state is fixed as $\ket{\Psi_6}$. This might be of interest for experimental implementations, as preparing a new state is arguably harder than adjusting the probing times.

Meanwhile, for the harmonic oscillator with varying $\tau$, the lowest-truncation state in the harmonic case is $\ket{\Psi_4}$ [Eq.~\eqref{eq:harmonic-optimal-n4-state}] at probing duration $\tau=1.774$ with $\mathbf{n}=4$. The same three scenarios as before are performed with the restriction $E_n\leq E_\mathbf{n}$ for $\mathbf{n}=4$ and the results are plotted in Fig.~\ref{fig:Kerr_n4}. The optimal state at $\alpha = 0.02$ is similar to the state $\ket{\Psi_4}$ at $\alpha=0$, while the state at $\alpha=-0.02$ is a ``flipped'' version of $\ket{\Psi_4}$: This is in the sense that the bulk of the negativity is concentrated into the region with $P_3^c = \frac{1}{3}$, while two peaks of positivity appear on either side in the regions with $P_3^c = \frac{2}{3}$.

\begin{figure}
    \includegraphics{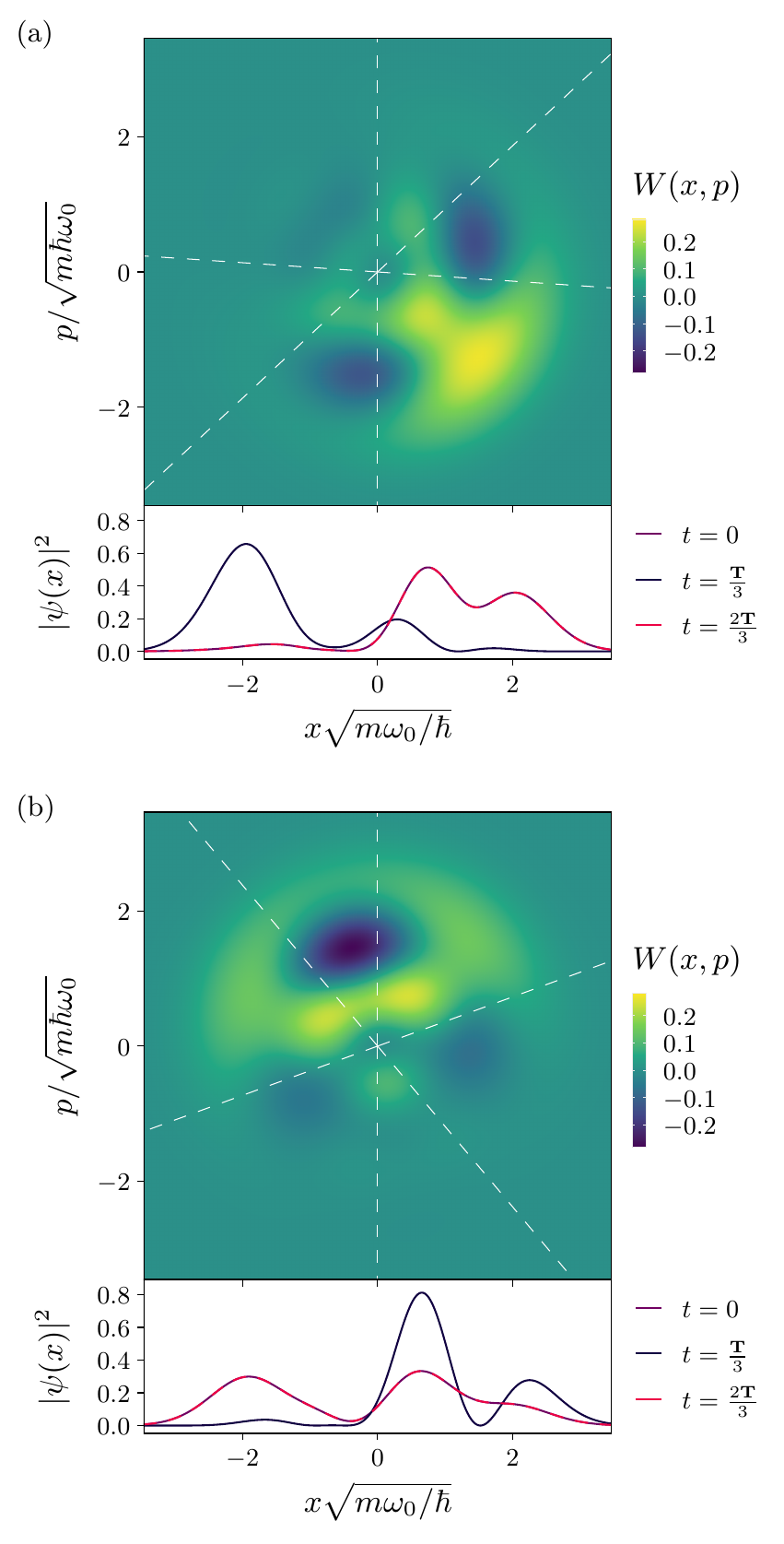}
    \caption{\label{fig:Kerr_n4_states}For the Kerr nonlinear system, the Wigner function and marginal probability distribution at times $t \in \{0,\mathbf{T}/3,2\mathbf{T}/3\}$ of the $\mathbf{n} = 4$ optimal state with (a) $\alpha=0.02$ and (b) $\alpha = -0.02$. The Wigner function is distributed such that there is a significant tail of the probability density that is present in the positive $x$ plane even when the bulk of the state is concentrated on the negative-$x$ plane.}
\end{figure}

The optimal probing duration similarly decreases with an increase in $\alpha$. However, unlike before, the quantum score increases with $\alpha$. We can understand this behavior as follows. The radial coordinate of the Wigner function corresponds to energy, which increases as we move further away from the origin. From Fig.~\ref{fig:Kerr_n4_states}(a) we find that the largest contribution to the optimal state comes from higher-energy states. The decrease in probing duration with the increase in anharmonicity ensures that the higher-energy contributions are in the positive-$x$ plane at $t=0$ and $2\mathbf{T}/3$. On the other hand, the period of the lower-energy states barely changes with anharmonicity, so the shorter probing duration means that these lower-energy contributions are able to remain in the positive-$x$ plane as the higher-energy contributions precess to the negative-$x$ plane at $t=\mathbf{T}/3$, which is present as the positive-$x$ tail in the marginal probability distribution shown in Fig.~\ref{fig:Kerr_n4_states}(a).

\subsection{Pendulum potential}\label{sec:example-pendulum}
The Hamiltonian of a physical pendulum is
\begin{equation}\label{eq:pendulum-hamiltonian}
\begin{aligned}
H(q=\phi,p=l) &= \frac{l^2}{2mb^2} - mb^2\omega_0^2\cos(\phi) \\
&= -\frac{4\alpha\omega_0 l^2}{\hbar} + \frac{\hbar\omega_0}{8\alpha}\cos(\phi),
\end{aligned}
\end{equation}
where $b$ is a length parameter, $\phi$ and $l$ are the angular coordinate and angular momentum of the system, respectively, and $\alpha:= - \hbar/8mb^2\omega_0 \leq 0$ is the anharmonicity. A small value of $\abs{\alpha} \ll 1$ allows for the usual approximation of the pendulum as a harmonic oscillator. Apart from describing physical pendulums and molecular interactions \cite{pendulum}, this Hamiltonian also describes a superconducting circuit without the small-angle approximation \cite{transmon}.

\subsubsection{Classical bounds}
A pendulum with energy $E$ has the period \cite{goldstein}
\begin{equation}\label{eq:pendulum-period}
T(E) = \frac{4}{\omega_0} K\pqty{\frac{8\abs{\alpha} E/\hbar\omega_0 + 1}{2}},
\end{equation}
where $K(m) = \int_0^{\pi/2}\dd{u}(1-m\sin^2u)^{-1/2}$ is the complete elliptic integral of the first kind.

The pendulum undergoes librations at large energies. However, the positions observable along the $x$ and $y$ directions commute, and it is known that commuting observables cannot exceed the classical score \cite{Zaw2022}, so no quantum violation of the protocol can be observed for librating states. Hence, we limit ourselves to closed nonlibrating trajectories, which limits us to the energy range $-1 \leq 8\abs{\alpha} E/\hbar\omega_0 < 1$.

Since the potential is even, we have $\Delta t_+(E) = \Delta t_-(E) = T_0(E)/2$ from Sec.~\ref{sec:simplify-probing-period}. The monotonicity of $K(m)$ within the energy range under consideration simplifies Eq.~\eqref{eq:energy-classical-condition} to
\begin{equation}\label{eq:pendulum-energy-bound}
\frac{3}{4}T_0(E_{\text{max}}) \leq \mathbf{T} \leq \frac{3}{2}T_0(E_{\text{min}})\,.
%\frac{3}{\omega_0}K\pqty{\frac{8\abs{\alpha} E_{\text{max}}/\hbar\omega_0 + 1}{2}} < \mathbf{T} < \frac{6}{\omega_0}K\pqty{\frac{8\abs{\alpha} E_{\text{min}}/\hbar\omega_0 + 1}{2}}.
\end{equation}
Although there is no closed-form expression for the inverse of $K(m)$, it can nonetheless be computed to an arbitrary precision. In terms of $K^{-1}(m)$, the energy bounds are
\begin{equation}
    \frac{2K^{-1}\pqty{\frac{\pi \tau}{3}}-1}{8\abs{\alpha}} \leq \frac{E_{\text{min}}}{\hbar\omega_0} \leq \frac{E}{\hbar\omega_0} \leq \frac{E_{\text{max}}}{\hbar\omega_0} \leq \frac{2K^{-1}\pqty{\frac{2\pi \tau}{3}}-1}{8\abs{\alpha}},
\end{equation}
where again $\mathbf{T}=:\tau\frac{2\pi}{\omega_0}$. The inclusion of the classical ground state requires $[2K^{-1}(\frac{\pi \tau}{3})-1]/8\abs{\alpha} \leq \hbar\omega_0/8\alpha \implies \tau \leq 3K(0)/\pi = \frac{3}{2}$. At the same time, the requirement $E_{\text{min}} \leq E_{\text{max}}$ implies $\tau \geq 3K(0)/2\pi = \frac{3}{4}$. Once again, we arrive at the restriction of the probing duration to the range $\frac{3}{4} \leq \tau \leq \frac{3}{2}$.

\subsubsection{Quantized system}
For the quantized system, the classical coordinates in the Hamiltonian are replaced by the observables $\Phi$ and $L$, which are defined as
\begin{equation}
    \Phi = \int_{-\pi}^{\pi}\dd{\phi} \phi \ketbra{\phi},\quad L = \sum_{m=-\infty}^\infty m\hbar \ketbra{m},
\end{equation}
with the resolution of the identity
\begin{equation}
    \int_{-\pi}^\pi\dd{\phi}\ketbra{\phi} = \sum_{m=-\infty}^\infty \ketbra{m} = \mathbbm{1}.
\end{equation}
We take the range $\phi \in (-\pi,\pi]$ for the angular coordinate. The angular wave function of the angular momentum eigenstate is $\braket{\phi}{m} = e^{im\phi}/\sqrt{2\pi}$, with which it can be verified that $\Phi$ and $L$ satisfy the Weyl relation ${e^{im\Phi}}{e^{i\phi L/\hbar}} = e^{-i m\phi} {e^{i\phi L/\hbar}}{e^{im\Phi}}$.

In the angular coordinate basis, the Schr\"odinger equation of the energy eigenstates reads
\begin{equation}
\pqty{4\alpha\frac{d^2}{d\phi^2} + \frac{1}{8\alpha}\cos(\phi)}\braket{\phi}{E_n} = \frac{E_n}{\hbar\omega_0}\braket{\phi}{E_n},
\end{equation}
which admits Mathieu functions that are $2\pi$ periodic in $\phi$ as solutions, with the parameter $-1/(4\alpha)^2$ and characteristic values $E_n/\hbar\omega_0\abs{\alpha}$ \cite{pendulum}. The eigenenergies are ordered $E_0 < E_1 < \cdots < E_n < \cdots$, with the associated wave function
\begin{equation}
    \braket{\phi}{E_n} = 
    \begin{cases}
    \frac{1}{\sqrt{\pi}}\operatorname{ce}_{n}\pqty{\phi/2,-1/(4\alpha)^2} & \text{for $n$ even} \\
    \frac{1}{\sqrt{\pi}}\operatorname{se}_{n+1}\pqty{\phi/2,-1/(4\alpha)^2} & \text{for $n$ odd,}
    \end{cases}
\end{equation}
where $\operatorname{ce}$ and $\operatorname{se}$ are the Mathieu functions. For brevity, we exclude the Mathieu function parameter $-1/(4\alpha)^2$ for the rest of this section.
\begin{widetext}
The matrix elements of $Q_3$ are found using Eq.~\eqref{eq:Wronskian-trick} to be
\begin{equation}
\mel{E_{n}}{Q_3}{E_{n'}}
=
\begin{cases}
\displaystyle\frac{
    \operatorname{ce}_{n}(\tfrac{\pi}{2})\operatorname{se}'_{n'+1}(\tfrac{\pi}{2}) -
    \operatorname{ce}_{n}(0)\operatorname{se}'_{n'+1}(0)
}{3\pi\pqty{E_n/\hbar\omega_0 - E_{n'}/\hbar\omega_0}}
2\alpha\pqty{
    1 + e^{i\tau\theta_{n,n'}} + e^{i2\tau\theta_{n,n'}}
} & \text{for $n$ even, $n'$ odd} \\[2ex]
\text{as above with $n \leftrightarrow n'$} & \text{for $n$ odd, $n'$ even} \\[1ex]
\displaystyle\frac{1}{2}\delta_{n,n'} & \text{otherwise.}
\end{cases}
\end{equation}
where $\theta_{n,n'} = 2\pi(E_n-E_{n'})/3\hbar\omega_0$ and $\operatorname{se}'(x) = \frac{d}{dx}\operatorname{se}(x)$. 
\end{widetext}

\subsubsection{Quantum violation}
\begin{figure}
    \includegraphics{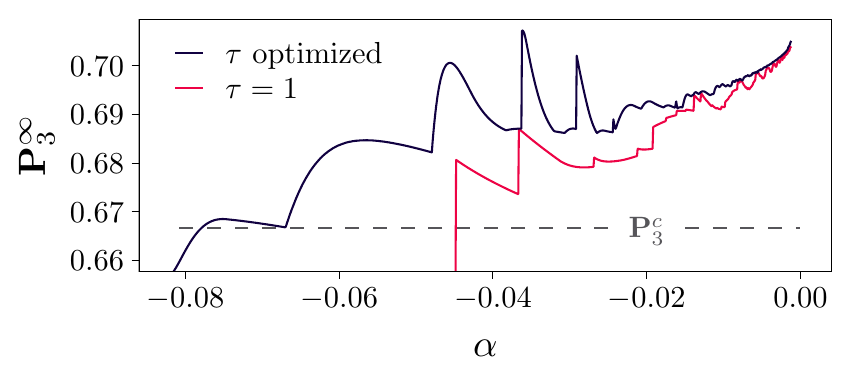}
    \caption{\label{fig:Pendulum_full_range}For the pendulum, the maximum quantum score against anharmonicity, for $\tau=1$ and the score maximized over $\frac{3}{4}\leq\tau\leq\frac{3}{2}$. The behavior is similar to that in Fig.~\ref{fig:Kerr_full_range}, although it deviates as the magnitude of the anharmonicity becomes large.}
\end{figure}

The maximum quantum score against anharmonicity is plotted in Fig.~\ref{fig:Pendulum_full_range} for both $\tau=1$ and the score maximized over $\frac{3}{4}\leq\tau\leq\frac{3}{2}$. The behavior is qualitatively similar to Fig.~\ref{fig:Kerr_full_range} for small anharmonicities, in that there is a general trend for the quantum score to decrease as the magnitude of the anharmonicity increases. However, for $\alpha < -0.02$, we find that larger scores, compared to the Kerr nonlinear system, can occur for the pendulum. This is expected, as the contribution of the higher-order terms of $\alpha$ can no longer be neglected as $\abs{\alpha}$ gets larger, resulting in more significant deviations between the two types of dynamics.

Another difference to note is that there is no longer a violation at $\alpha < -0.08$ for the pendulum, while the Kerr nonlinear system can still beat the classical bound at $\alpha \sim -0.1$. This is largely due to the energy bounds: For the same value of anharmonicity, the energy bounds in Eq.~\eqref{eq:pendulum-energy-bound} result in a much smaller subspace for the pendulum potential than the truncation for the Kerr nonlinear system.

\subsubsection{Comparison with the harmonic approximation}\label{sec:pendulum-weak-regime}
Upon expanding the quantized version of Eq.~\eqref{eq:pendulum-hamiltonian} up to the fourth order in $\Phi$, defining the annihilation operator $a := \Phi/\sqrt{8\abs{\alpha}} + i \sqrt{8\abs{\alpha}}L/\hbar$, and performing the rotating-wave approximation \cite{transmon}, we recover Eq.~\eqref{eq:kerr-quantum-hamiltonian}. The anharmonicity defined for the pendulum corresponds exactly to the $\alpha$ in Eq.~\eqref{eq:kerr-quantum-hamiltonian}. Here the anharmonicity is always nonpositive, with a weaker anharmonicity corresponding to a larger amplitude of the potential.

The comparison with the harmonic approximation performed in Sec.~\ref{sec:kerr-weak-regime} for the Kerr nonlinearity is repeated for the pendulum potential in Appendix~\ref{apd:pendulum-weak-regime}. For the values of anharmonicity considered, the results agree with what was obtained for the Kerr nonlinear systems, and the optimal states are similar in both cases. However, we notice that the quantum scores are slightly smaller for the pendulum than the Kerr nonlinear system at $\alpha \sim -0.02$. This is as we observed with the full range of anharmonicities in Fig.~\ref{fig:Kerr_full_range}, where the quantum scores begin to diverge between the two cases as $\alpha \lesssim -0.02$.

\subsection{\label{sec:example-morse}Morse potential}
So far, the systems we have studied are even and their trajectories are closed orbits. Our protocol is general enough that these are not conditions necessary for the procedure to work, as we will encounter in this section.

The Morse potential is used to model molecular interactions, which also takes into consideration the effects of bond breaking \cite{morse-quantum}. The Hamiltonian with the Morse potential is given by
\begin{equation}
\begin{aligned}
    H(x,p) &= \frac{p^2}{2m} + D_e\pqty{1-e^{cx}}^2,
\end{aligned}
\end{equation}
where $D_e$ is the dissociation energy and $c$ is a length scale. Note that the potential as defined here differs from the usual convention in two ways: The origin is placed at the equilibrium position $x_0$ so that $x_0 = 0$ and the coordinates have been flipped from $x \to -x$ so that the particle escapes to $x\to-\infty$ upon dissociation. This was done to put Eq.~\eqref{eq:P3-definition} in its standard form. The same physics would be obtained by keeping the usual convention for the potential and using $\Pr[x(t) < x_0]$ in place of $\Pr[x(t) > 0]$ in Eq.~\eqref{eq:P3-definition}.

\subsubsection{Classical bounds}
Classically, the solution for $E < D_e$ with the initial condition $p(t=0) = 0$ is \cite{morse-classical}
\begin{equation}\label{eq:morse-classical-bound-state}
x(t) = \frac{1}{c}\log{\frac{1-E/D_e}{1-\sqrt{E/D_e}\cos[{2\pi t}/{T_0(E)}]}},
\end{equation}
where the period $T_0(E)$ is given by
\begin{equation}
T_0(E) = \frac{2\pi/\omega_0}{\sqrt{1-E/D_e}},
\end{equation}
where we have defined the natural frequency $\omega_0 := \sqrt{2D_ec^2/m}$. An expression similar to Eq.~\eqref{eq:morse-classical-bound-state} exists for $E > D_e$. However, in that case, the particle escapes to $x\to-\infty$, so the period is taken to be $T_0(E)=\infty$. Hence, the times that the particle remains in the positive and negative regions are, respectively,
\begin{align}
\Delta t_-(E) &= \begin{cases}
\frac{2\pi-2\acos\sqrt{E/D_e}}{\omega_0\sqrt{1-E/D_e}} & \text{for $E < D_e$} \\
\infty & \text{for $E \geq D_e$,} 
\end{cases}\\
\Delta t_+(E) &= \begin{cases}
\frac{2\acos\sqrt{E/D_e}}{\omega_0\sqrt{1-E/D_e}} & \text{for $E < D_e$} \\ 
\frac{2}{\omega_0} & \text{for $E = D_e$} \\ 
\frac{2\operatorname{arccosh}\!\sqrt{E/D_e}}{\omega_0\sqrt{E/D_e-1}} & \text{for $E > D_e$.}
\end{cases}
\end{align}
Therefore, $\min_E \Delta t_-(E) = \pi/\omega_0$ and $\max_E \Delta t_+(E) = \pi/\omega_0$ for all possible values of $E$. Notice how $\Delta t_\pm(E)$ is exactly the same as for the harmonic oscillator. Therefore, for a system with the Morse potential, we can perform the original protocol without modification.

\subsubsection{Quantized system}
In the quantum case, the bound eigenstates are exactly solvable \cite{morse-quantum}, with the energy eigenvalues
\begin{equation}
E_n = \hbar\omega_0\pqty{n+\frac{1}{2}}\pqty{1-\frac{1}{2\lambda}\pqty{n+\frac{1}{2}}}
\end{equation}
for $n=0,1,\dots,\lfloor \lambda - \frac{1}{2} \rfloor$, where we have defined $\lambda:=2D_e/\hbar\omega_0$ for notational convenience. The corresponding wave functions and their derivatives are \cite{morse-ladder}
\begin{align}
\braket{x}{E_n} &= \sqrt{ \frac{cn!\pqty{2\lambda-2n-1}}{\pqty{2\lambda-n-1}!} }
\pqty{2\lambda e^{cx}}^{\lambda-n-1/2} e^{-\lambda e^{cx}} \\[-2ex]
&\quad{}\times{} L_n^{(2\lambda-2n-1)}\!\pqty{2\lambda e^{cx}},\nonumber \\[1.5ex] \label{eq:morse-eigenstate-derivative}
\frac{d}{dx}\braket{x}{E_n} &= 
c\bqty{\pqty{\lambda-n-\frac{1}{2}} - \frac{\lambda^2 e^{cx}}{\lambda-n}}\braket{x}{E_n} \\[0ex]
&\quad{}-{} \frac{c\lambda e^{cx}}{\lambda-n}\sqrt{\frac{n(2\lambda-n)(\lambda-n-\frac{1}{2})}{\lambda-n+\frac{1}{2}}} \braket{x}{E_{n-1}}.\nonumber
\end{align}
Here $L_n^{(\alpha)}\!(z) = \sum_{k=0}^n (-1)^k\spmqty{n+\alpha\\n-k}{z^k}/{k!}$ are the generalized Laguerre polynomials.
\begin{widetext}
Placing Eq.~\eqref{eq:morse-eigenstate-derivative} into Eq.~\eqref{eq:Wronskian-trick}, the matrix elements of $\sgn(X)$ for $n \neq n'$ can be found to be
\begin{equation}
\begin{aligned}
\mel{E_{n}}{\sgn(X)}{E_{n'}}
= \frac{2}{c\pqty{n-n'}\pqty{2\lambda-(1+n+n')}}\Bigg\{
    &-(n-n')\pqty{1 + \frac{\lambda^2}{(\lambda-n)(\lambda-n')}}\braket{E_n}{0_x}\!\braket{0_x}{E_{n'}} \\
    &\quad{}-{}
    \frac{\lambda}{\lambda-n}\sqrt{\frac{n(2\lambda-n)(\lambda-n-\frac{1}{2})}{\lambda-n+\frac{1}{2}}}\braket{E_{n-1}}{0_x}\!\braket{0_x}{E_{n'}} \\
    &\quad{}+{}
    \frac{\lambda}{\lambda-n'}\sqrt{\frac{n'(2\lambda-n')(\lambda-n'-\frac{1}{2})}{\lambda-n'+\frac{1}{2}}}\braket{E_n}{0_x}\!\braket{0_x}{E_{n'-1}}
\Bigg\},
\end{aligned}
\end{equation}
where $\ket{0_x} = \ket{x=0}$. Meanwhile, the diagonal elements are
\begin{equation}\label{eq:morse-diagonal-matrix-element}
\mel{E_n}{\sgn(X)}{E_n} =
\frac{4\pqty{2\lambda}^{2\lambda-2n-1}e^{-2\lambda} P^{n}(\lambda)}{(2\lambda - n-1)!}\pqty{2\lambda - 2n - 1} + 
2Q(2\lambda-2n-1,2\lambda) - 1,%\qquad\text{for $n \leq 6$}
\end{equation}
where $Q(a,x)=\int_x^\infty\dd{t} t^{a-1} e^{-t}/(a-1)!$ is the regularized $\Gamma$ function and $P^{n}(\lambda)$ is a polynomial of degree $n-1$ given in Eq.~\eqref{eq:morse-polynomials}.
\end{widetext}

The coefficients of $P^{n}(\lambda)$ grow large with $n$ very quickly, which results in some difficulty in computing higher-order matrix elements of $\sgn(X)$ due to the numerical instability. Hence, we consider only the comparison with the harmonic approximation performed in the weak-anharmonicity and low-energy regime for the Morse potential. For this reason, we will also ignore the free-particle solutions within the continuous spectrum of the Hamiltonian that only occur for large energies with $E>D_e$.

\subsubsection{Comparison with the harmonic approximation}
We repeat the study performed in Sec.~\ref{sec:kerr-weak-regime} for the Morse potential in Appendix~\ref{apd:morse-weak-regime}. While we again find that the quantum score approaches the harmonic case as $\alpha\to0$, the same score requires anharmonicities that are an order of magnitude smaller here compared to the previous cases. This is due to the shape of the Morse potential, where $\lim_{x\to-\infty} V(x) = D_e$ is finite in comparison to $\lim_{x\to\infty} V(x) = \infty$, so the particle is much less effectively trapped from the negative-$x$ direction. Therefore, the Morse potential needs to be much deeper, hence less anharmonic, to confine the particle as effectively as the other potentials.

\subsection{Particle in an infinite well\label{sec:example-infinite-well}}
Although the particle in an infinite well exhibits pathological behaviors in both the classical (phase-space trajectories are not continuous) and quantum (momentum operator is not self-adjoint) cases, this toy model serves to illustrate the generality of the protocol. We show how the procedure laid out in Sec.~\ref{sec:summary-procedure} can be applied to this system, even when the particle in an infinite well is not approximately harmonic at small energies.

Its Hamiltonian is usually written as
\begin{equation}
H(x,p) = \frac{p^2}{2m} + V(x),\qquad V(x) = \begin{cases}
0 & \text{if $\abs{x} \leq L/2$} \\
\infty & \text{otherwise.}
\end{cases}
\end{equation}
The quantized Hamiltonian is obtained with a direct replacement $x\to X$ and $p\to P$. In the classical and quantum cases, the trajectories and wave functions, respectively, have support only within the range $x \in [-L/2,L/2]$, where $x$ is the position and $L$ is the length of the well.

\subsubsection{Classical bounds}
Apart from the fixed points $(x(t),p(t)) = (x(0),0)$, every classical trajectory is closed and passes the position $x=-L/2$. Since every other classical state can be obtained by an offset $t \to t-t_0$, without any loss of generality, we only need to consider the trajectories with $x(0) = -L/2$. These trajectories are uniquely specified by the energy $E = p^2/2m$, and their position in time is
\begin{equation}
    x(t) = -\frac{L}{2} + \sgn\bqty{ \sin(\frac{2\pi t}{T_0(E)}) }
    \sqrt{\frac{m}{2E}}\pqty{t\bmod\frac{T_0(E)}{2}},
\end{equation}
where the period of oscillation is
\begin{equation}
    T_0(E) = \sqrt{\frac{2m}{E}}L =: \frac{2\pi/\omega_0}{\sqrt{E/\hbar\omega_0}}.
\end{equation}
For later convenience, we define the natural frequency of the system as $\omega_0 := 2\pi^2\hbar/mL^2$.

To find $\Delta_\pm(E)$, we notice that the potential $V(x)$ is even, so Sec.~\ref{sec:simplify-probing-period} implies
\begin{equation}
    \Delta t_+(E) = \Delta t_-(E) = \frac{2\pi/\omega_0}{2\sqrt{E/\hbar\omega_0}}.
\end{equation}
When performing the protocol with $\mathbf{T} =: \tau\frac{2\pi}{\omega_0}$, the energy bounds that satisfy the condition in Eq.~\eqref{eq:energy-classical-condition} for $\tau$ are
\begin{equation}\label{eq:infinite-well-conditions}
    \frac{9}{16\tau^2} \leq \frac{E_{\text{min}}}{\hbar\omega_0} \leq \frac{E}{\hbar\omega_0} \leq \frac{E_{\text{max}}}{\hbar\omega_0} \leq \frac{9}{4\tau^2}.
\end{equation}

\begin{figure}
    \includegraphics{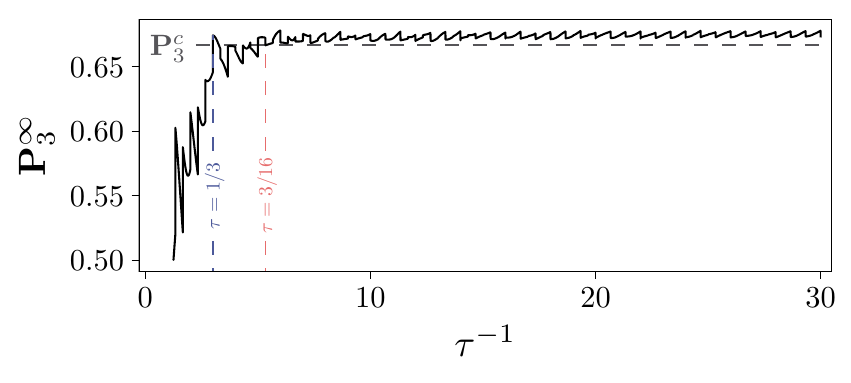}
    \caption{\label{fig:infinite-well-full-range}For the infinite well, plot of $\mathbf{P}_3^\infty$ against $\tau^{-1}$. There is no violation for all values of $\tau > \frac{1}{3}$, while all values of $\tau < \frac{3}{16}$ violate the classical bound.}
\end{figure}

\subsubsection{Quantized system}
The solutions for the quantum particle in an infinite well are well known to be
\begin{equation}
E_n = \pqty{\frac{n}{2}}^2\hbar\omega_0,
\quad
\braket{x}{E_n} = \begin{cases}
\sqrt{\frac{2}{L}}\cos(\frac{n\pi}{L}x) & \text{$n$ odd}\\
\sqrt{\frac{2}{L}}\sin(\frac{n\pi}{L}x) & \text{$n$ even,}
\end{cases}
\end{equation}
where $n\geq1$ and $\omega_0$ is as previously defined. Hence, we can obtain the matrix elements for $Q_3$ using Eq.~\eqref{eq:Wronskian-trick},
\begin{widetext}
\begin{equation}
\mel{E_n}{Q_3}{E_{n'}} = \begin{cases}
\displaystyle\frac{1}{2}\delta_{n,n'} & \text{if $(n-n')\bmod 2 = 0$} \\[1ex]
\frac{
    1 + 
    e^{i(n^2-n^{\prime 2})({\pi\tau}/{6})} +
    e^{i(n^2-n^{\prime 2})({2\pi\tau}/{6})}
}{3}\frac{1}{\pi}\pqty{\frac{1}{n+n'} + \frac{(-1)^{n\bmod 2}}{n-n'}} & \text{otherwise.}
\end{cases}
\end{equation}
\end{widetext}
Meanwhile, rewriting Eq.~\eqref{eq:infinite-well-conditions} in terms of $n$,
\begin{equation}\label{eq:infinite-well-truncation}
\frac{3}{2\tau} < n < \frac{3}{\tau}.
\end{equation}
Therefore, we can construct the matrix for $Q_3$ explicitly with the truncation given in Eq.~\eqref{eq:infinite-well-truncation} to find $\mathbf{P}_3^\infty$.

\begin{figure*}
    \includegraphics{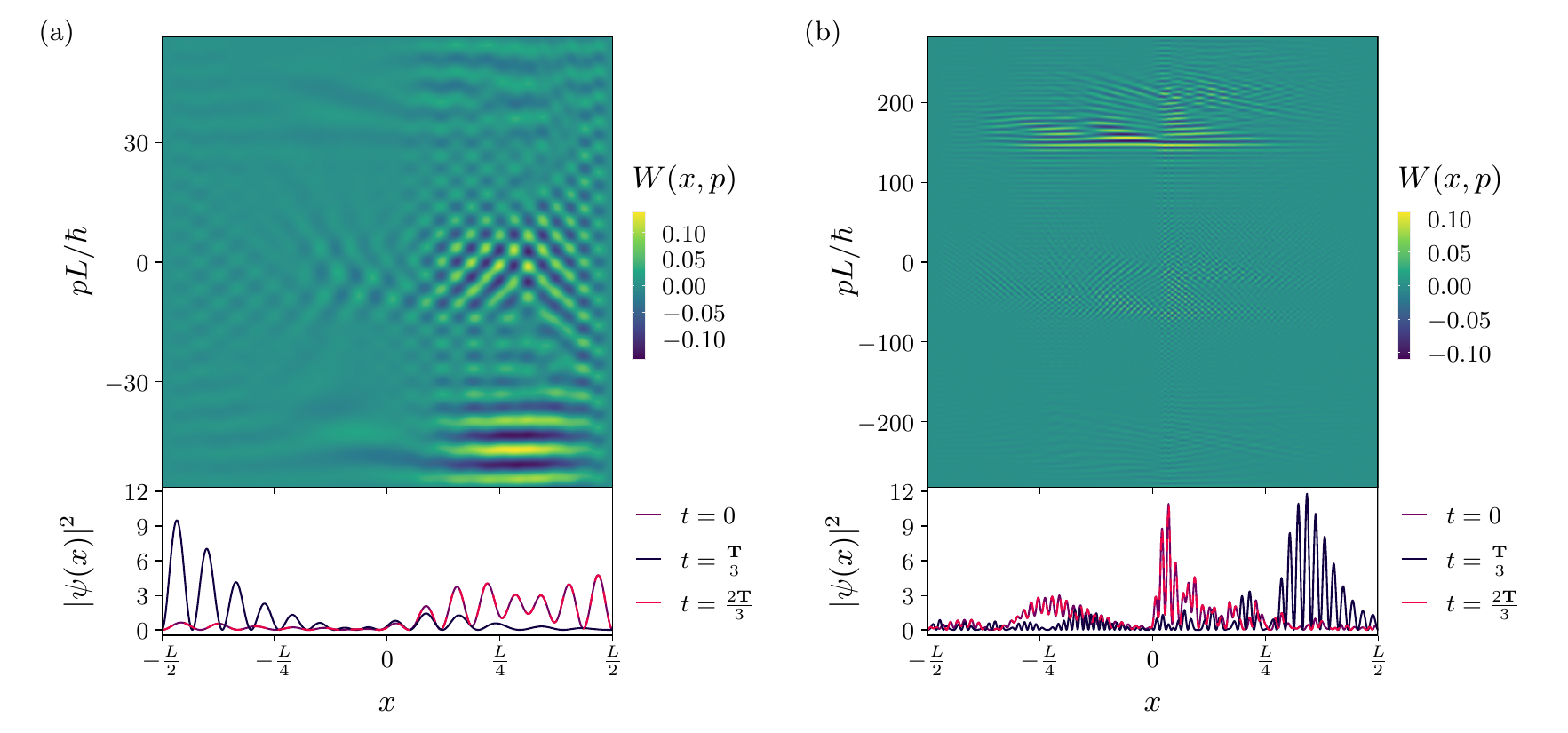}
    \caption{\label{fig:infinite-well-probabilities}For the infinite well, Wigner functions and probability densities of the optimal state for (a) $\tau=\frac{1}{3}$ and (b) $\tau = \frac{1}{30}$ at the three different probing times. The interference patterns of the Wigner function in the latter case are intricate (they can be seen more clearly in the probability distributions), but the working principle is similar to the former case. The probability densities at times $0$ and $2\mathbf{T}/3$ overlap in both cases. Just like in Fig.~\ref{fig:Harmonic_n4_states}, the quantum score is augmented by the positive-$x$ tail when the state is mostly in the negative-$x$ axis.}
\end{figure*}

\subsubsection{Quantum violation}
The maximum quantum score $\mathbf{P}_3^\infty$ is plotted against the inverse of the probing duration $\tau^{-1}$ in Fig.~\ref{fig:infinite-well-full-range}. Two particular regions of $\tau$ are also highlighted: There is no violation for all values of $\tau > \frac{1}{3}$, while all values of $0 < \tau < \frac{3}{16}$ violate the classical bound.

The probability distributions of the optimal states are plotted in Fig.~\ref{fig:infinite-well-probabilities} for $\tau=\frac{1}{3}$ and $\tau=\frac{1}{30}$. Their behavior is reminiscent of Fig.~\ref{fig:Harmonic_n4_states}: A violation of the classical bound is achieved by having a small but consequential contribution from the positive-$x$ tail of the state when most of the state is concentrated on the negative-$x$ axis.

Similar states known as ultraslow states have been studied in the context of the quantum projectile \cite{Rocket}. Equation~\eqref{eq:infinite-well-conditions} constrains the speed of the particle so that it should only travel a distance of $L/2 \leq \Delta x \leq L$ between successive probing times. However, the positive-$x$ tail shows that there is a nonzero probability of the particle traveling much slower and remaining in the positive-$x$ axis (hence ultraslow), increasing the quantum score.

\section{Conclusion}

Dynamics-based certification of quantumness is a witness of nonclassicality for continuous-variable systems, based on the assumption that the dynamics is known. It was previously introduced for the harmonic oscillator \cite{tsirelson,Zaw2022}. In this paper we have extended the protocol to time-independent Hamiltonians. We have characterized the generic conditions under which a classical-quantum gap is expected and have explicitly worked out four examples: an anharmonic oscillator, a pendulum (where one of the conjugate variables is compact and therefore the other has a discrete spectrum), an asymmetric potential with open orbits, and a system that cannot be approximated as harmonic even at low energies. While we have placed our focus on the certification of nonclassicality in the context of quantum theory, we note that the protocol and its classical bounds depend only on measuring position, bounding energy, and the requirement that the Hamiltonian is the generator of time evolution. So the study could carry over to more general theories that uphold those features \cite{general-Hamiltonian}, possibly with a different degree of violation of the classical bound.

With an eye on experimental implementations, we have shown that the original harmonic protocol is robust to weak anharmonicity in the low-energy regime; simple modifications like adjusting the probing time can already provide significant improvements. For example, transmon qubits are engineered to be weakly anharmonic with $\abs{\alpha} \ll 1$ \cite{transmon} and the state space $\{\ket{n}\}_{n=0}^\infty$ is restricted to $n \ll n_{\text{crit}}$ for some critical photon number in the dispersive regime \cite{critical-transmon}, which are exactly the conditions we have considered. Similar conditions and approximations are also found in a wide variety of optical systems, ranging from nonlinear fibers \cite{Kerr-fiber} to optomechanics \cite{Kerr-optomechanics}.

A possible generalization to this protocol would be to use probing times $\{0,t_1,t_2\}$ that are not equally spaced, as considered in the Tsirelson paper \cite{tsirelson} for the harmonic case. This requires updating the classical condition in Eq.~\eqref{eq:energy-classical-condition} to $\max_{E} \Delta t_+(E) \leq t_2$ and $t_1 \leq \min_{E}\Delta t_-(E)$. The rest of the protocol proceeds analogously. One could also use more than three probing times, although this may not be beneficial: For uniform precessions, this was found to decrease the classical-quantum gap \cite{Zaw2022}.

Additionally, the protocol readily extends itself to multiple degrees of freedom. Equations~\eqref{eq:delta-t-neg} and \eqref{eq:delta-t-pos} make reference only to the measured observable $q$, and would more generally quantify how long the state remains in the positive or negative regions of $q$ in a higher-dimensional phase space. Since that is enough to specify the classical score, the condition in Eq.~\eqref{eq:general-classical-condition} maintains $\mathbf{P}_3^c = \frac{2}{3}$. Therefore, barring the simplifications at the end of the section that are specific to systems with a single degree of freedom, the procedure laid out in Sec.~\ref{sec:general} can still be used for systems with multiple degrees of freedom.

In this paper we have taken the dynamics of the system as given and studied the protocol for what we believe to be a reasonable choice of the generalized coordinate $q$. This determines the states, whose nonclassicality can be detected. Conversely, one's goal may be to detect the nonclassicality of a given state: Then one would look for the coordinate $q$ for a given $H$, with which the classical bound can be violated for that state, or even better optimize $H$ for that task. In particular, it remains an open question whether initial states with a positive Wigner function can violate the classical bound for some dynamics. For Hamiltonians that are at most quadratic in position and momentum, like a free particle or the harmonic oscillator, the time evolutions of both the classical probability distribution in phase space and the Wigner function are exactly the same, which means that a negative Wigner function is necessary for a violation of the classical bound. In those cases, this protocol acts as a negativity witness for the state. For a general time-independent Hamiltonian, while the canonical variables have the same time evolution for both the classical and quantum cases, the states might not. As such, it is not clear if Wigner negativity is still a necessary condition for a violation of the classical bound or if some Hamiltonians enable dynamics-based certification of quantumness of Wigner positive states or even Gaussian states.

\section*{Acknowledgments}
This research was supported by the National Research Foundation, Singapore and A*STAR under its CQT Bridging Grant.

\bibliography{refs}

\appendix

\section{Comparisons to the harmonic approximation}
We repeat the study of the protocol in the comparison to the harmonic approximation, reported in Sec.~\ref{sec:kerr-weak-regime} for the Kerr-nonlinear system, with both the pendulum (Sec.~\ref{sec:example-pendulum}) and Morse potentials (Sec.~\ref{sec:example-morse}). We consider low-energy states of the form $\sum_{n\leq\mathbf{n}}\psi_n \ket{E_n}$ and weak anharmonicities with $\abs{\alpha} \ll 1$.

For both systems, the Hamiltonian approaches that of the harmonic oscillator as $\alpha \to 0$. At this limit,
\begin{equation}\label{eq:harmonic-limit-optimal-n6-state}
    \ket{\Psi_6} = \frac{1}{\sqrt{2}}\pqty{
        \frac{4}{\sqrt{21}} \ket{E_0} - \ket{E_3} + \sqrt{\frac{5}{21}}\ket{E_6}
    }
\end{equation}
is the optimal state that achieves the largest quantum score $P_3 = 0.687$ for $n\leq \mathbf{n}=6$ and $\tau=1$, while
\begin{equation}\label{eq:harmonic-limit-optimal-n4-state}
\begin{aligned}
    \ket{\Psi_4} &= 
        \sqrt{0.279}\ket{E_0} +
        \sqrt{0.191}\ket{E_1}e^{-i\theta_4} +
        \sqrt{0.121}\ket{E_2}e^{-i2\theta_4} \\
        &\quad {}+{}\sqrt{0.309}\ket{3}e^{-i3\theta_4} +
        \sqrt{0.100}\ket{E_4}e^{-i4\theta_4},
\end{aligned}
\end{equation}
with $\theta_4 = 0.215\pi$, achieves the largest quantum score $P_3 = 0.669$ for $n\leq \mathbf{n}=4$ and $\tau=1.774$.

In this low-energy regime, we study the following three scenarios.

(i) Under the truncation $\mathbf{n} = 6$ ($\mathbf{n} = 4$), we find the optimal state $\ket{\Psi}$ and time $\tau$ that violates the classical bound maximally. This fully takes into account the anharmonicity of the system.

(ii) We use the state $\ket{\Psi_6}$ ($\ket{\Psi_4}$) and find the optimal $\tau$. This only corrects for the change in the period of the state caused by the anharmonicity.

(iii) We use the state $\ket{\Psi_6}$ and $\tau = 1$ ($\ket{\Psi_4}$ and $\tau=1.774$). This ignores the anharmonicity of the system completely.

\subsection{Pendulum potential}\label{apd:pendulum-weak-regime}
\begin{figure}
    \includegraphics{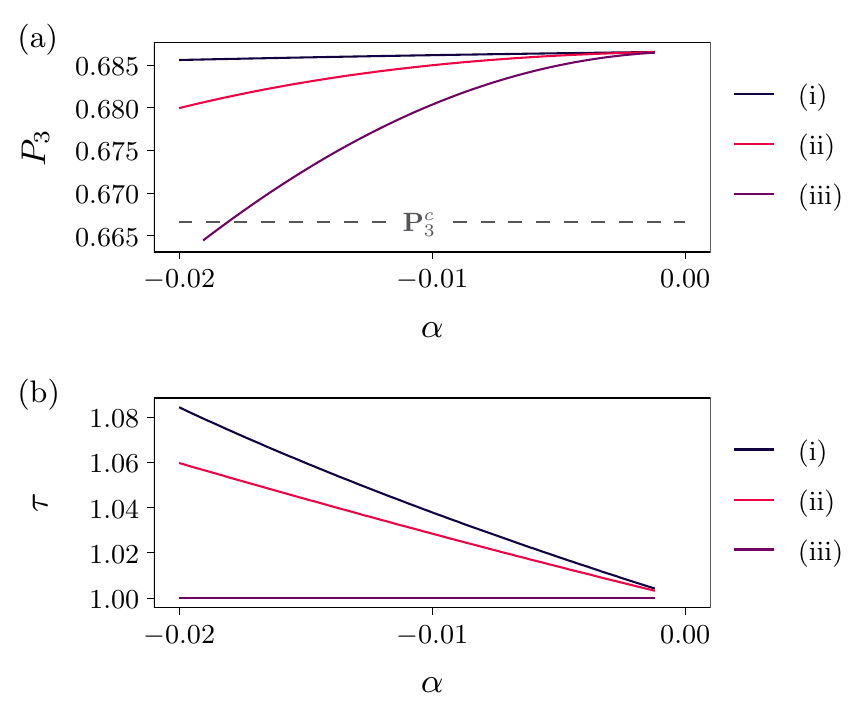}
    \caption{\label{fig:Pendulum_n6}Same as Fig.~\ref{fig:Kerr_n6} but for the pendulum, for low-energy states with truncation $\mathbf{n}=6$. (a) Score $P_3$ as a function of the anharmonicity $\alpha$ for the three scenarios described above. (b) Optimal $\tau$ used for scenarios (i) and (ii).}
\end{figure}

\begin{figure}
    \includegraphics{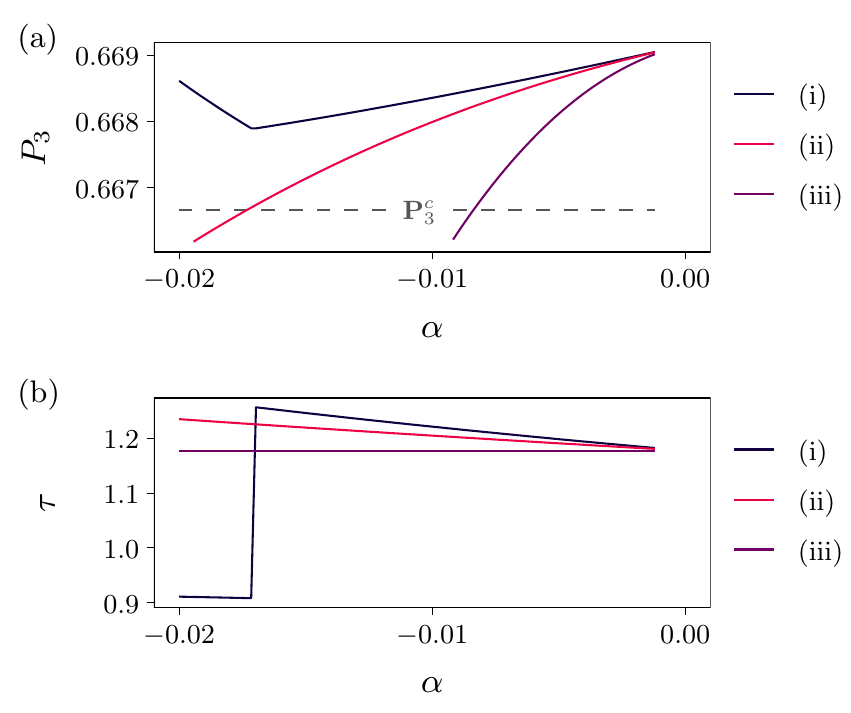}
    \caption{Same as Fig.~\ref{fig:Kerr_n4} but for the pendulum, for low-energy states with truncation $\mathbf{n}=4$. (a) Score $P_3$, optimized over $\tau$, plotted against $\alpha$ (b) Optimal $\tau$ that achieves the optimized quantum score.}
    \label{fig:Pendulum_n4}
\end{figure}

\begin{figure*}
    \includegraphics{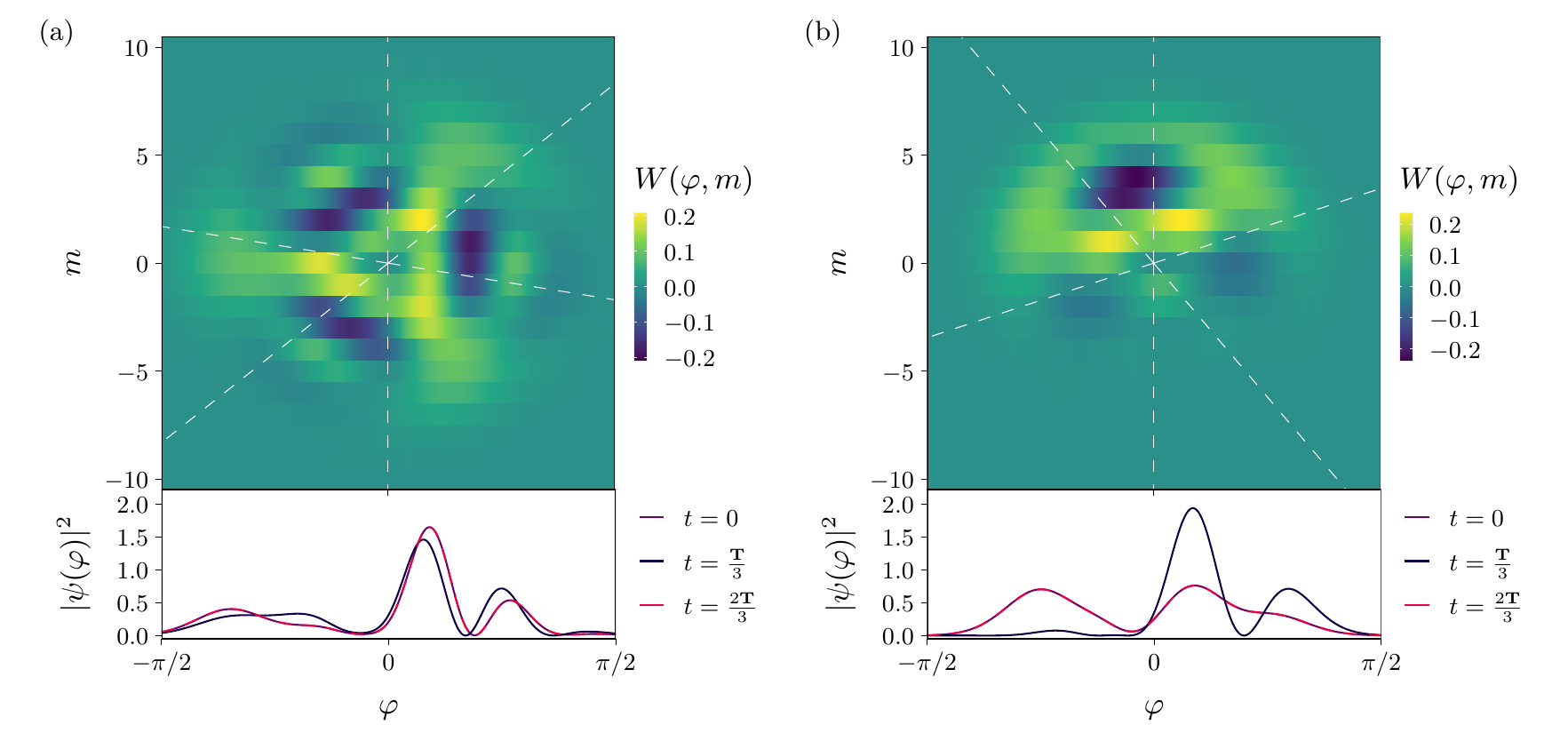}
    \caption{\label{fig:Pendulum_states}For the pendulum, the angular Wigner function $W(\varphi,m)$ \cite{angular-momentum-phase-space} and marginal probability density of the optimal low-energy state at $\alpha=-0.02$ with truncation (a) $\mathbf{n}=6$ and (b) $\mathbf{n}=4$. Note that the angular momentum $m\hbar$ only takes discrete values with integer $m$. These states have the same features as Figs.~\ref{fig:Kerr_n4_states} and \ref{fig:Kerr_n6_states}.}
\end{figure*}

For the pendulum potential, we plot the three scenarios with the low-energy states with truncations $\mathbf{n}=6$ (Fig.~\ref{fig:Pendulum_n6}) and $\mathbf{n}=4$ (Fig.~\ref{fig:Pendulum_n4}). As mentioned in the main text, the results here agree with what was previously obtained, although they begin to deviate as $\alpha \sim -0.02$. The optimal states obtained are also similar, as illustrated with the Wigner functions plotted in Fig.~\ref{fig:Pendulum_states}.

\subsection{Morse potential}\label{apd:morse-weak-regime}
\begin{figure}
    \includegraphics{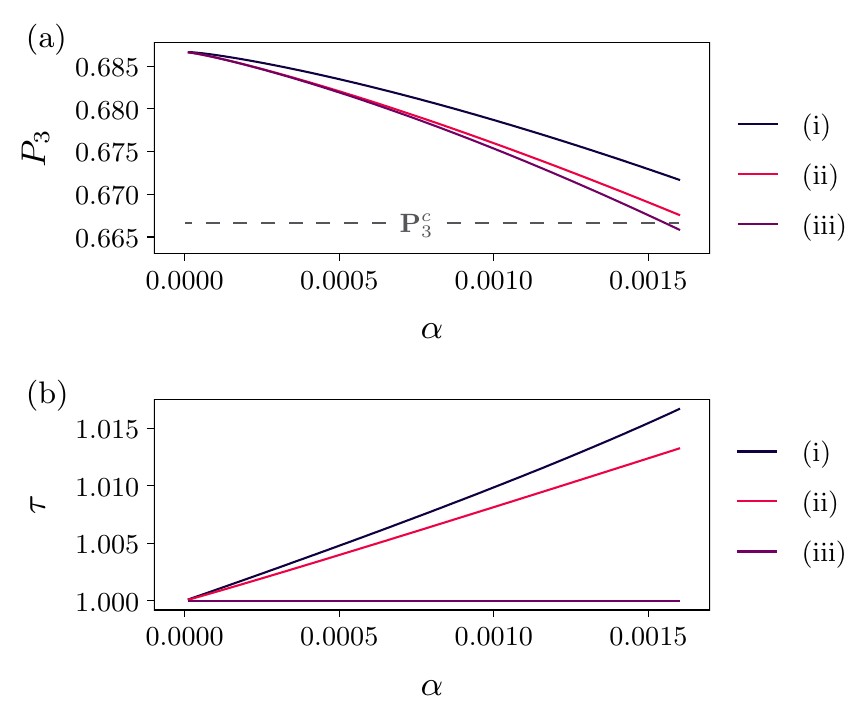}
    \caption{Same as Fig.~\ref{fig:Kerr_n6} but for the Morse potential, for low-energy states with truncation $\mathbf{n} = 6$. (a) Score $P_3$ as a function of $\alpha$ for the three scenarios. (b) Optimal $\tau$. Note that $\alpha$ is an order of magnitude smaller than in Figs.~\ref{fig:Kerr_n6} and \ref{fig:Pendulum_n6}.}
    \label{fig:Morse_n6}
\end{figure}

\begin{figure}
    \includegraphics{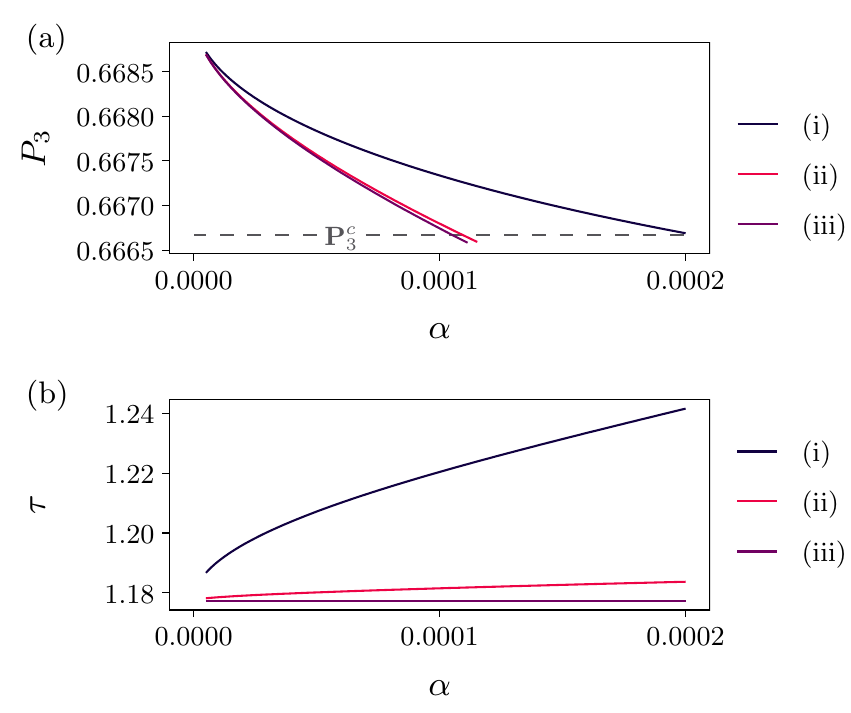}
    \caption{Same as Fig.~\ref{fig:Kerr_n4} but for the Morse potential, for low-energy states with truncation $\mathbf{n} = 4$. (a) Score $P_3$, optimized over $\tau$, is plotted against $\alpha$. (b) Optimal $\tau$. Note that $\alpha$ is about two orders of magnitude smaller than in Figs.~\ref{fig:Kerr_n4} and \ref{fig:Pendulum_n4}.}
    \label{fig:Morse_n4}
\end{figure}

\begin{figure*}
    \includegraphics{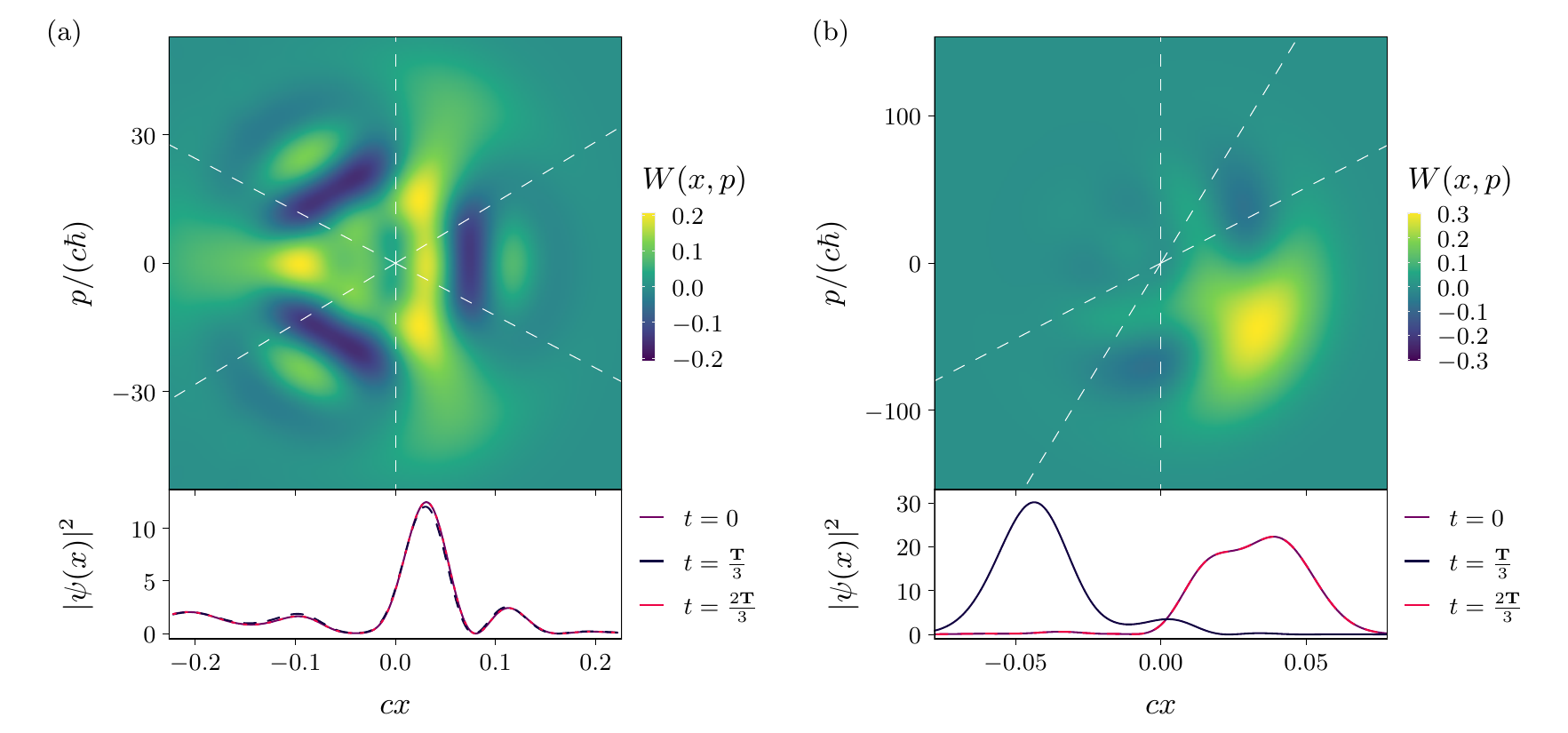}
    \caption{\label{fig:Morse_states}For the Morse potential, the Wigner function $W(x,p)$ and marginal probability density of the optimal low-energy state at (a) $\mathbf{n} = 6$ and $\alpha=1.6\times10^{-3}$ and (b) $\mathbf{n} = 4$ and $\alpha=2\times10^{-4}$. These states have the same features as in Figs.~\ref{fig:Kerr_n4_states} and \ref{fig:Kerr_n6_states}, but there is far less negativity present.}
\end{figure*}

The results for the Morse potential with truncation $\mathbf{n}=6$ is plotted in Fig.~\ref{fig:Morse_n6}. Note that the horizontal axis is an order of magnitude smaller than those in Figs.~\ref{fig:Kerr_n6} and \ref{fig:Pendulum_n6}, so a much weaker anharmonicity is required in the Morse potential for a violation of the classical bound. For the truncation $\mathbf{n}=4$, shown in Fig.~\ref{fig:Morse_n4}, the anharmonicities are two orders of magnitude smaller than those in Figs.~\ref{fig:Kerr_n4} and \ref{fig:Pendulum_n4}. The Wigner functions of the optimal state for both truncations $\mathbf{n}=6$ and $\mathbf{n}=4$ are shown in Fig.~\ref{fig:Morse_states}.

\section{\label{apd:morse-diagonal}Diagonal Matrix Elements for the Morse potential}
The diagonal matrix elements of $\sgn(X)$ for the Morse potential is given by
\begin{equation}
\begin{aligned}
&\mel{E_n}{\sgn(X)}{E_n} \\
&\quad{}={} \int_{0}^{\infty}\dd{x}\abs{ \braket{x}{E_n} }^2 - \int_{-\infty}^0\dd{x}\abs{ \braket{x}{E_n} }^2 \\
&\quad{}={} 1 - 2\frac{n!\pqty{2\lambda-2n-1}}{\pqty{2\lambda-n-1}!} \\
&\qquad\quad{}\times{}\int_{0}^{2\lambda}
 \dd{z}
z^{2\lambda-2n-2} e^{-z} \pqty{L_n^{(2\lambda-2n-1)}\!(z)}^2,
\end{aligned}
\end{equation}
where the integration variable was changed to $z=2\lambda e^{cx}$ and $L_n^{(\alpha)}\!(z) = \sum_{k=0}^n (-1)^k\spmqty{n+\alpha\\n-k}{z^k}/{k!}$ are the generalized Laguerre polynomials. The integral can be worked out by using the explicit form of Laguerre polynomials and the definition of the regularized $\Gamma$ function
\begin{equation}
\begin{aligned}
&\frac{1}{(2\lambda-2n-2+k)!}\int_{0}^{2\lambda}\dd{z} z^{2\lambda-2n-2+k} e^{-z}\\
&\qquad{}={} 1 - Q(2\lambda-2n-1+k,2\lambda),
\end{aligned}
\end{equation}
which satisfy the recursive relation
\begin{equation}
    Q(\alpha+k,2\lambda) = Q(\alpha+(k-1),2\lambda) + \frac{(2\lambda)^{\alpha+(k-1)}}{[\alpha+(k-1)]!}e^{-2\lambda},
\end{equation}
with which the above integral can be simplified to the form given in Eq.~\eqref{eq:morse-diagonal-matrix-element}. The first few polynomials $P^{n}(\lambda)$ are given in Eq.~\eqref{eq:morse-polynomials}. However, because the size of the coefficients grows large very quickly, we were only able to compute the matrix elements of $\sgn(X)$ for the weak anharmonicity and low-energy regime:
\begin{widetext}
\begin{subequations}\label{eq:morse-polynomials}
\begin{align}
P^0(\lambda) &= 0, \\
P^1(\lambda) &= 1, \\
P^2(\lambda) &= 6 + 2\lambda, \\
P^3(\lambda) &= 180 - 42\lambda + \frac{32}{2}\lambda^2, \\
P^4(\lambda) &= 6440 - \frac{5828}{3}\lambda + 190\lambda^2 + \frac{58}{3}\lambda^3, \\
P^5(\lambda) &= 347\,760 - 140\,604\lambda + \frac{102\,376}{5}\lambda^2 - \frac{4912}{5}\lambda^3 + \frac{212}{3}\lambda^4,\\
P^6(\lambda) &= 32\,802 - \frac{450\,763}{30}\lambda + \frac{372\,931}{150}\lambda^2 - \frac{215\,779}{1350}\lambda^3 + \frac{31\,411}{8100}\lambda^4 + \frac{19}{108}\lambda^5,\\
P^7(\lambda) &= 1\,979\,385\,408 - \frac{4\,965\,563\,888}{5}\lambda + \frac{6\,608\,820\,632}{35}\lambda^2 - \frac{5\,199\,530\,008}{315}\lambda^3 + \frac{73\,504\,376}{105}\lambda^4 - \frac{553\,792}{45}\lambda^5 + \frac{1192}{3}\lambda^6.
\end{align}
\end{subequations}
\end{widetext}

\end{document}